# A Taxonomy of Data Grids for Distributed Data Sharing, Management and Processing

Srikumar Venugopal, Rajkumar Buyya* and Kotagiri Ramamohanarao
Grid Computing and Distributed Systems Laboratory,
Department of Computer Science and Software Engineering,
The University of Melbourne, Australia
Email:{srikumar, raj, rao}@cs.mu.oz.au

**Abstract**

Data Grids have been adopted as the platform for scientific communities that need to share, access, transport, process and manage large data collections distributed worldwide. They combine high-end computing technologies with high-performance networking and wide-area storage management techniques. In this paper, we discuss the key concepts behind Data Grids and compare them with other data sharing and distribution paradigms such as content delivery networks, peer-to-peer networks and distributed databases. We then provide comprehensive taxonomies that cover various aspects of architecture, data transportation, data replication and resource allocation and scheduling. Finally, we map the proposed taxonomy to various Data Grid systems not only to validate the taxonomy but also to identify areas for future exploration.

Through this taxonomy, we aim to categorise existing systems to better understand their goals and their methodology. This would help evaluate their applicability for solving similar problems. This taxonomy also provides a "gap analysis" of this area through which researchers can potentially identify new issues for investigation. Finally, we hope that the proposed taxonomy and mapping also helps to provide an easy way for new practitioners to understand this complex area of research.

## 1   Introduction

The next-generation of scientific applications in domains as diverse as high-energy physics, molecular modeling and data mining involve the production of large datasets from simulations or from large-scale experiments. These datasets have to be shared among large groups of researchers spread worldwide and their analysis is highly compute-intensive requiring dedicated resources. Collectively, these large scale applications have come to be known as part of e-Science [1], a discipline that envisages using high-end computing, storage, networking and Web technologies together to facilitate collaborative, data-intensive scientific research[1]. However, this requires new paradigms in Internet computing that address issues such as multi-domain applications, co-operation and co-ordination of resource owners and blurring of system boundaries. Grid computing [2] is one such paradigm that proposes aggregating geographically-distributed, heterogeneous computing, storage and network resources to form Grids that provide unified, secure and pervasive access to the combined capabilities of the aforementioned resources.

Data Grids [3][4][5] are Grids where the access to distributed data resources and their management are treated as first-class entities along with processing operations. Data Grids, therefore primarily deal with providing services and infrastructure for distributed data-intensive applications. The fundamental features of Data Grids are provision of a secure, high-performance transfer protocol for transferring large datasets and a scalable replication mechanism for ensuring distribution of data on-demand. However, in order to enable researchers to derive maximum benefits out of the infrastructure, the following are needed: (a) ability to search through numerous available datasets for the required dataset, (b) ability to discover suitable data resources for accessing the data and computational resources for performing analysis, (c) ability to select suitable computational resources and process data on them and (d) ability for resource owners to manage access permissions. Thus, seamless organisation, well-defined architecture and intelligent resource allocation and scheduling are also required to ensure that users realise their utility from the Data Grid infrastructure.

---

*Contact Author

[1] also known as e-Research with the inclusion of digital libraries and the humanities community.

The explosion in popularity of Data Grids in scientific and commercial settings has led to a variety of systems offering solutions for dealing with distributed data-intensive applications. Unfortunately, this has also led to difficulty in evaluating these solutions because of the confusion in pinpointing their exact target areas. Also, there exist a few different mechanisms with similar properties for supporting a distributed data-intensive infrastructure. They are content delivery networks, peer-to-peer networks and distributed databases. The main objective of this paper is to, therefore, delineate very clearly the uniqueness of Data Grids through systematic characterisation and differentiation from other similar paradigms and through the taxonomy, provide a basis against which present and future developments in this area can be categorised. As a result, the taxonomy proposed in this paper provides readers with a detailed understanding of Data Grid technologies and techniques and helps them identify important and outstanding issues for further investigation.

A few studies have investigated and surveyed Grid research in the recent past. In [6], the authors present a taxonomy of various Grid resource management systems that focuses on the general resource management architectures and scheduling policies. Specifically for Data Grids, Bunn and Newman provide an extensive survey of projects in High Energy Physics in [7] while Qin and Jiang [8] produce a compilation that concentrates more on the constituent technologies. In [9], authors identify functional requirements (features and capabilities) and components of a persistent archival system. In contrast to these papers, Finkelstein, et al [10] spell out requirements for Data Grids from a software engineering perspective and elaborate on the impact that these have on architectural choices. A similar characterisation has been performed by Mattmann, et al [11]. Our work in this paper however concentrates on issues specific to all data-intensive application environments including Data Grids. Also, we expand our scope to include data transport and replication along with resource management and scheduling within Data Grids. The taxonomy proposed in this paper is exhaustive and complements the earlier surveys by providing a more detailed and complete understanding of Data Grids and its underlying technologies through multiple perspectives including software engineering, resource allocation, data management and user requirements.

The rest of this paper is organised as follows. Section 2 presents a general overview of Data Grids, its characteristics and compares it to other distributed data-oriented network technologies. Next, section 3 provides taxonomies for Data Grids, data transport technologies, data replication mechanisms and resource allocation and scheduling policies. Section 4 then surveys some representative projects and research works and classifies them according to the taxonomies provided. Finally, a summary is provided and the paper is concluded.

## 2 Overview

Figure 1 shows a high-level view of a Data Grid consisting of storage resources that are connected by high speed networks spanning continents. The thick lines show high bandwidth networks linking the major centres and the thinner lines are lower capacity networks that connect the latter to their subsidiary centres. The data generated from an instrument, experiment or a network of sensors is stored in its principal storage site and is transferred to the other storage sites around the world on request. A replication mechanism creates and manages the copies at various locations. The replication mechanism is guided by replication strategies that take into account current and future demand for the datasets, locality of requests and storage capacity to create copies of the data. Users within an area connect to their local repository to obtain the data they require if they have been granted the requisite rights and permissions. If the data is not present, then it is fetched from a remote repository. The data maybe transmitted to a computational site such as a cluster or a supercomputer facility for processing. After processing, the results may be sent to a visualisation facility, a shared repository or to the desktops of the individual users.

Another aspect of a Data Grid is to maintain shared collections of data distributed across administrative domains. These collections are maintained independent of the underlying storage systems and are able to in-clude new sites without major effort. More importantly, it is required that the data and information associated with data such as metadata, access controls and version changes be preserved even in the face of technology changes. These requirements lead to the establishment of persistent archival storage [12].

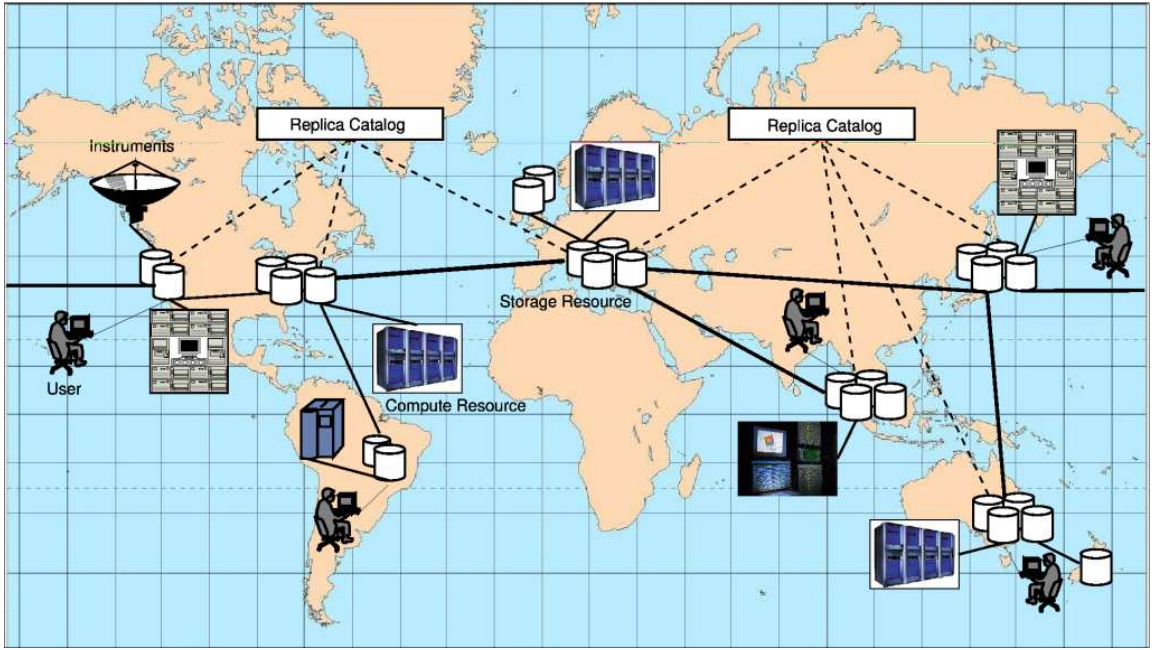

Figure 1: A High-Level view of Data Grid.

A Data Grid , therefore, provides a platform through which users can access aggregated computational, storage and networking resources to execute their data-intensive applications on remote data. It promotes a rich environment for users to analyse data, to share the results with their collaborators and to maintain state information about the data seamlessly across institutional and geographical boundaries. Often cited examples for Data Grids are the ones being set up for analysing the huge amounts of data that will be generated by the CMS, ATLAS, Alice and LHCb experiments at the Large Hadron Collider (LHC) [13] at CERN when they will begin production in 2007. These Data Grids will involve thousands of physicists spread over hundreds of institutions worldwide and will be replicating and analysing terabytes of data daily.

The presence of a large number of users belonging to different organisations and sharing common resources exhibits its own unique set of characteristics detailed below:

- *Proliferation of Data:* Data-intensive applications are characterised by the presence of large datasets in the order of Gigabytes(GB) and beyond. For example, the CMS experiment at the LHC is expected to produce 1 PB ($10^{15}$ bytes) of RAW data and 2 PB of event summary data (ESD) annually when it begins production [14].

- *Geographical Distribution:* Present-day scientific collaborations involve researchers from different countries across continents. Data has to be made available to all the collaborators thus requiring an infrastructure spanning continents.

- *Single Source:* More often than not, the principal instrument such as a telescope or a particle accelerator is the single source of data generation for these experiments. This means that all data is written at a single point and then replicated for read access. Thus, consistency requirements are limited to ensuring that the copy of data is the exact mirror of that at the source. Updates to the source are propogated to the replicas either by the replication mechanism or by a separate consistency management service.

- *Unified Namespace:* The data in a Data Grid share the same logical namespace in which every data element has a unique logical filename. The logical filename is mapped to one or more physical files on various storage resources across the Grid. The logical namespace is managed by a distributed catalog mechanism.

- *Limited resources:* Resources such as clusters, supercomputers, storage nodes and high-speed network bandwidth are all shared between the various groups involved. Even though the aggregation of these resources represents higher power than each of them individually, there is a need for intelligent resource management to ensure effective and fair share of the resources for everyone.

- *Local Autonomy:* While various organisations contribute their resources to a Data Grid, they still retain control over the resources. In particular, they can decide who can access the resources. Therefore, within a collaboration, there is a potential for different levels of access rights among users.

- *Access Restrictions:* It follows from the previous property that usage of resources is dependent on if a user is granted access and the level of access so provided. Frequently, users might wish to ensure confidentiality of their data or restrict distribution to close collaborators. Thus, mechanisms for ensuring security of data need to be in place in a Data Grid which is much harder than a simple centralised data repository.

- *Heterogeneity:* Data Grid environments encompass various hardware and software configurations that potentially use different protocols. Applications that have to work between different systems, such as copying data from a tape archival system in one organisation to a disk drive in another, would have to be engineered to work across multiple interfaces. The existence of a standard protocol that supports querying, transfer and archiving of data among different systems simplifies development of Data Grid applications. This protocol would also allow the application to determine the access rights available to the user on whose behalf it is performing the task.

Thus it can be seen that due to the intensive processing, transfer and storage requirements, a situation could arise where many users are contending for share of a few resources. Resource owners being independent could set terms of access and security requirements leaving users and applications to deal with numerous and potentially incompatible infrastructures. Unpredictable variations in resource availability leads to the infrastructure being too unreliable for production-level usage. To tackle these challenges, Foster, Kesselman and Tuecke [15] have proposed a Grid architecture for resource sharing among different entities based around the concept of *Virtual Organizations(VOs)*. A VO is formed when different organisations come together to share resources and collaborate in order to achieve a common goal. A VO defines the resources available for the participants and the rules for accessing and using the resources. Resources here are not just compute, storage or network resources, they may also be software, scientific instruments or business data. A VO mandates the existence of a common middleware platform that provides secure and transparent access to common resources. A VO also provides tools and mechanisms for applications to determine the suitability and accessibility of available resources. In practical terms, a VO may be created using mechanisms such as Certificate Authorities(CAs) and trust chains for security, replica management systems for data organisation and retrieval and centralised scheduling mechanisms for resource management.

## 2.1 Layered Architecture

The components of a Data Grid can be organised in a layered architecture as shown in Figure 2. This architecture follows from similar definitions in [15]and in [16]. Each layer builds on the services offered by the lower layer in addition to interacting and co-operating with components and the same level (eg. Resource broker invoking VO services). We can describe the layers as:

- **Grid Fabric**: Consists of the distributed computational resources (clusters, supercomputers), storage resources (RAID arrays, tape archives) and instruments (telescope, accelerators) connected by high-bandwidth networks. Each of the resources runs system software such as operating systems, job submission and management systems and relational database management systems (RDBMS).

- **Communication**: Consists of protocols used to transfer data between the resources in the Grid Fabric layer. These protocols are built on core communication protocols such as TCP/IP and authentication protocols such as PKI (Public Key Infrastructure), passwords or SSL (Secure Sockets Layer). The cryptographic protocols allow verification of users' identity and ensure security and integrity of transferred data. File transfer protocols such as vanilla FTP or GSIFTP provide services for efficient transfer of data between two resources on the Data Grid.

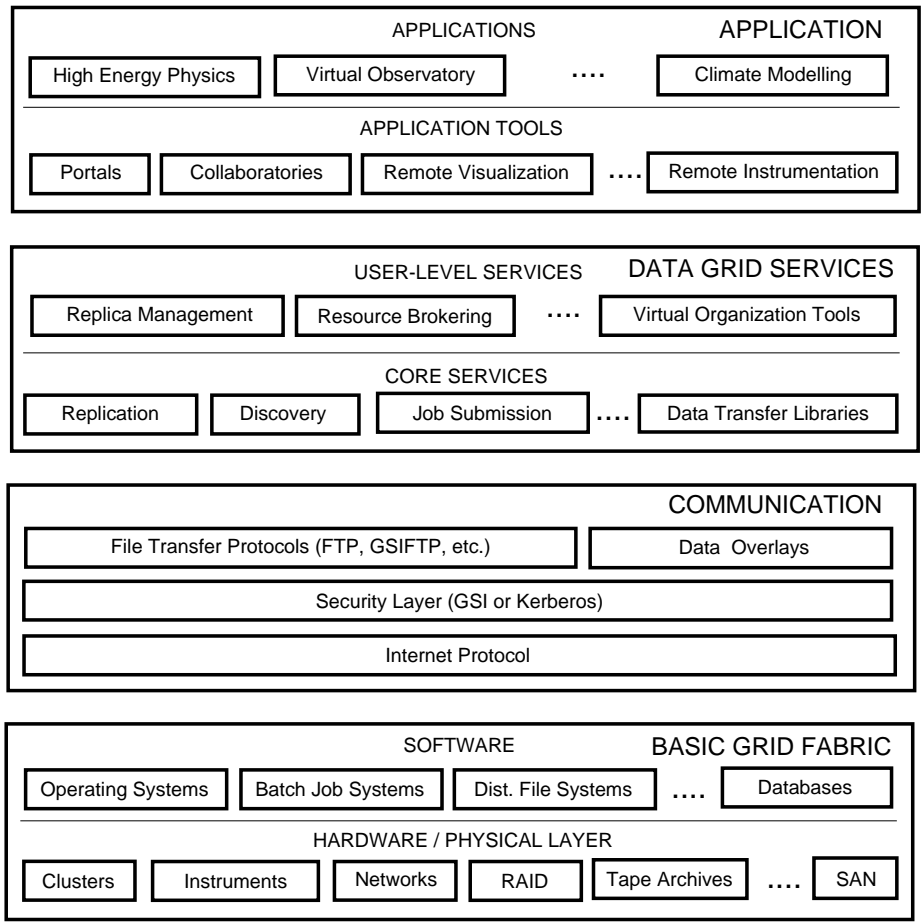

Figure 2: A Layered Architecture.

- **Data Grid Services**: Provides services for managing, transferring and processing data in a Data Grid. The core level services such as replication, data discovery and job submission provide access transparency to distributed data and computation. User-level services such as resource brokering and replica management provide global mechanisms that allow for efficient resource management hidden behind simple interfaces and APIs. VO tools provide easy way to perform functions such as adding new resources to a VO, querying the existing resources and for managing users' access rights.

- **Applications**: Tools such as portals and collaboratories provide domain-specific services to users by calling on services provided by the layers below. Each domain provides a familiar interface and access to services such as visualisation.

The services layer is collectively known as Grid middleware. The middleware provides a straightforward way for applications to invoke the services provided by itself and the lower layers while abstracting out much of the inherent complexity and heterogeneity.

## 2.2   Related Data-Intensive Research Paradigms

Three related distributed data-intensive research areas that share similar requirements, functions and characteristics are described below. These have been chosen because of the similar properties and requirements that they share with Data Grids.

### 2.2.1 Content Delivery Network

A Content Delivery Network(CDN) [17, 18] consists of a "collection of (non-origin) servers that attempt to offload work from origin servers by delivering content on their behalf" [19]. That is, within a CDN, client requests are satisfied from other servers distributed around the Internet (edge servers) that cache the content originally stored at the source(origin) server. A client request is rerouted from the main server to an available server closest to the client likely to host the content required [18]. This is done by providing a DNS(Domain Name System) server that resolves the client DNS request to the appropriate edge server. If the latter does not have the requested object then it retrieves the data from the origin server or another edge server. The primary aims of a CDN are therefore, load balancing to reduce effects of sudden surges in requests, bandwidth conservation for objects such as media clips, large images, etc. and reducing the round-trip time to serve the content to the client. CDNs are generally employed by Web content providers and companies such as Akamai [20], Speedera [21] and IntelliDNS [22] are commercial providers who have built dedicated infrastructure to serve multiple clients. However, CDNs haven't gained wide acceptance for data distribution because of the restricted model that they follow. Also, current CDN infrastructures are proprietary in nature and owned completely by the providers.

### 2.2.2 Peer-to-Peer Network

Peer-to-peer(P2P) networks [23][24] are formed by ad hoc aggregation of resources to form a decentralised system within which each peer is autonomous and depends on other peers for resources, information and forwarding requests. The primary aims of a P2P network are: to ensure scalability and reliability by removing the centralised authority, to ensure redundancy, to share resources and to ensure anonymity. An entity in a P2P network can join or leave anytime and therefore, algorithms and strategies have to be designed keeping in mind the volatility and requirements for scalability and reliability. P2P networks have been designed and implemented for many target areas such as compute resource sharing (e.g. SETI@Home [25], Compute Power Market [26]), content and file sharing (Napster [27], Gnutella [28], Kazaa [29]) and collaborative applications such as instant messengers (Jabber [30]). A detailed taxonomy and survey of peer-to-peer systems can be found in [31]. Here we are concerned mostly with content and file-sharing P2P networks as these involve data distribution. Such networks have mainly focused on creating efficient strategies to locate particular files within a group of peers, to provide reliable transfers of such files in the face of high volatility and to manage high load caused due to demand for highly popular files. Currently, none of the major P2P content sharing networks support computation in addition to data distribution. However, efforts are on to provide such services.

### 2.2.3 Distributed Databases

A distributed database(DDB) [32, 33] is a logically organised collection of data stored at different sites of a computer network such that each site has a degree of autonomy, is capable of executing a local application and also, participates in the execution of a global application. A distributed database can be formed either by taking an existing single site database and splitting it over different sites (top-down approach) or by federating existing database management systems so that they can be accessed through a uniform interface (bottom-up approach) [34]. Varying degrees of autonomy are provided ranging from tightly-coupled sites to complete site independence. Distributed databases have evolved to serve the needs of large organisations which need to remove the need for a centralised computer centre, to interconnect existing databases, to replicate databases to increase reliability and to add new databases as new organisational units are added. This technology is very robust. It provides distributed transaction processing, distributed query optimisation and efficient management of resources. However, these systems cannot be employed in their current form at the scale of Data Grids envisioned as they have strict requirements for ACID (Atomicity, Consistency, Isolation and Durability) properties[35] to ensure that the state of the database remains consistent and deterministic.

## 2.3 Analysis of Data-intensive Networks

This section compares the data-intensive paradigms described in the previous sections with Data Grids in order to bring out the uniqueness of the latter by highlight the respective similarities and differences. Also, each of these areas have their own mature solutions which may be applicable to the same problems in Data Grids either wholly or with some modification based on the differing properties of the latter. These properties are summarised in Table 1 and are explained as below:

**Purpose** - Considering the purpose of the network, it is generally seen that P2P content sharing networks are vertically integrated solutions for a single goal (file-sharing) while Data Grids support various activities such as analysis and production of data and collaboration over the same infrastructure. CDNs are dedicated to caching web content so that clients are able to access it faster. DDBs are used for integrating existing diverse databases to provide a uniform, consistent interface for querying and/or for replicating existing databases for increasing reliability or throughput.

**Aggregation** - Aggregation is the way in which resources are brought together for creating the network and can be *specific*, when the resource providers have some sort of agreement with each other and *ad hoc*, when a resource can enter or leave the network at will.

**Organisation** - The organisation of a CDN is hierarchical with the data flowing from the origin to the edges. Data is cached at the various edge servers to exploit locality of data requests. There are many models for organisation of P2P content sharing network and these are linked to the searching methods for files within the network. Within Napster, a peer has to connect to a centralised server and search for an available peer that has the required file. The two peers then directly communicate with each other. Gnutella avoids the centralised directory by having a peer broadcast its request to its neighbours and so on until the peer with the required file is obtained. Kazaa and FastTrack limit the fan-out in Gnutella by restricting broadcasts to SuperPeers who index a group of peers. Freenet [36] uses content-based hashing, in which a file is assigned a hash based on its contents and nearest neighbour search is used to identify the required document. Thus, three different models of organisation, viz. centralised, two-level hierarchy and flat (structured and unstructured) can be seen in the examples presented above. Distributed databases provide a relational database management interface and so, are organised accordingly. Global relations are split into fragments that are allocated to either one or many physical sites. In the latter case, replication of fragments is carried out to ensure reliability of the database. While distribution transparency may be achieved within top-down databases, it may not be the case with federated databases that have varying degrees of heterogeneity and autonomy.

Table 1: Comparison between various data distribution networks

| Property | Data Grids | P2P (Content sharing) | CDN | DDB |
|---|---|---|---|---|
| Purpose | Analysis, collaboration | File sharing | reducing web latency | Integrating existing databases, Replicating database for reliability & throughput |
| Aggregation | Specific | Ad hoc | Specific | Specific |
| Organisation | Hierarchical, federation | Centralised, two-level hierarchy, flat | Hierarchical | Centralised, federation |
| Access Type | Mostly read with rare writes | Mostly read with frequent writes | Read-only | Equally read and write |
| Data Discovery | Replica Catalog | Central directory, Flooded requests or document routing | HTTP Request | Relational Schemas |
| Replication | Locality or Popularity, One or more primary copies | Popularity, No primary copy | Caching, Primary copy | Fragment replication, Primary copy |
| Consistency | Weak | Weak | Strong (read-only) | Strong |
| Transaction Support | None currently | None | None currently | Yes |
| Processing Requirements | Data Production and Analysis | None currently | None (Client-side) | Query Processing |
| Autonomy | Access, Operational, Participation | Operational, Participation | None (Dedicated) | Operational (federated) |
| Heterogeneity | Hardware, System, Protocol, Representation | Hardware, System, Representation | Hardware, Representation | Hardware, System (federated), Representation |
| Management Entity | VO | Individual | Single Organisation | Single Organisation |
| Security Requirements | Authentication, Authorisation, Data Integrity | Anonymity | Data Integrity | Authentication, Authorisation, Data Integrity |
| Latency Management & Performance | Replication, Caching, Streaming, Pre-staging, Optimal selection of data and processing resources | Replication, Caching, Streaming | Caching, Streaming | Replication, Caching |

**Access Type**  - Access type distinguishes the type of data access operations conducted within the network. Data Grids and P2P networks are mostly read-only environments with the number of writes varying with the number of entities uploading or modifying data. CDNs are almost exclusively read-only environments for end-users and updating of data happens at the origin servers only. In DDBs, data is both read and written frequently.

**Data Discovery**  - Another distinguishing property is how the data is discovered within the network. The role of replica catalogues for searching for data within Data Grids has been mentioned before. The three approaches for searching within P2P networks have been mentioned previously. Current research focuses on the document routing model and the four algorithms proposed for this model: Chord [37], CAN [38], Pastry [39] and Tapestry [40]. CDNs fetch data which has been requested by a browser through HTTP (Hyper Text Transfer Protocol). DDBs are organised using the same relational schema paradigm as single-site databases and thus, data can be retrieved using SQL(Structured Query Language).

**Replication**  - Replication categorises the manner in which copies of data are created and maintained in the network. Replication has twin objectives: one, as will be seen later, to increase performance by reducing latency and the other, to provide reliability by creating multiple backup copies of data. In a Data Grid, data is mostly replicated on the basis of the locality of requests especially based on community-driven hierarchical model. In P2P networks such as Gnutella and Freenet, replicas are created proportional to the popularity of the original datasets [41]. While several replication strategies have been suggested for a CDN, Karlsson and Mahalingam [42] experimentally prove that replication provides worse performance than caching. Within DDBs, replication is at the fragment level and there are varying degrees of replication from none to partial to full replication [43]. Replication can also be distinguished on the basis of whether there is a primary copy, changes of which causes updates to other replicas. While data in CDNs and DDBs have primary copies,

P2P content distribution networks generally have no notion of a primary copy due to their ad-hoc nature and the absence of strict ownership controls. In Data Grids with hierarchical organisation, there is a single data source which act as priamry copy (e.g., LHC Grid). In those with federation organisation, have more than one data sources that act as primary copies for their respective datasets (e.g., eDiaMoND [44] and BIRN [45]). However, it is upto the discretion of the replica sites to subscribe to the updates to the primary copies.

**Consistency** - Consistency is an important property which determines how "fresh" the data is. Grids and P2P networks generally do not provide strong consistency guarantees because of the overhead of maintaining locks on huge volumes of data and the ad hoc nature of the network respectively. Among the exceptions for Data Grids is the work of Dullman, et al. [46] which discusses a consistency service for replication in Data Grids. In P2P networks, Oceanstore [47] is a distributed file system that provides strong consistency guarantees through expensive locking protocols. In CDNs, while the data in a cache may go stale, the system always presents the latest version of the data when the user requests it. Therefore, the consistency provided by a CDN is strong.

Distributed databases, as mentioned before, have strong requirements for satisfying ACID properties. Therefore, semantics for updating are much more stricter within distributed databases than in other distribution networks. Also, updates are more frequent and can happen from within any site in the network. These updates have to be migrated to other sites in the network so that all the copies of the data are synchronised. There are two methods for updating that are followed [48]: *lazy*, in which the updates are asynchronously propagated and *eager*, in which the copies are synchronously updated.

**Transaction Support** - A transaction is a set of operations (actions) such that all of them succeed or none of them succeed. Transaction support implies the existence of check-pointing and rollback mechanisms so that a database or data repository can be returned to its previous consistent state in case of failure. It follows from the discussion of the previous property that transaction support is essential for distributed databases. CDNs have no requirements for transaction support as they only support read only access to data to the end users. P2P Networks and Data Grids currently do not have support for recovery and rollback. However, efforts are on to provide transaction support within Data Grids to provide fault tolerance for distributed transactions [49] [50].

**Processing Requirements** - Processing of data implies that the data is transferred to a node to be an input to another process or program that evaluates it and returns the results to the user. As mentioned before, analysis of data is integrated within Data Grids. CDNs are exclusively data-oriented environments with a client accessing data from remote nodes and processing it at its own site. While current P2P content sharing networks have no processing of the data, it is possible to integrate such requirements in the future. DDBs involve query processing which can be held at a designated site that has the master copy of the data (replicated databases) or at any site (federated or multi-database systems).

**Autonomy** - Autonomy deals with the degree of independence allowed to different nodes within a network. However, there could be different types and different levels of autonomy provided [34, 51]. *Access autonomy* allows a site or a node to decide whether to grant access to a user or another node within the network. *Operational autonomy* refers to the ability of a node to conduct its own operations without being overridden by external operations of the network. *Participation autonomy* implies that a node has the ability to decide the proportion of resources it donates to the network and the time it wants to associate or disassociate from the network. Data Grid nodes have all the three kinds of autonomy to the fullest extent. While nodes in a P2P network do not have fine-grained access controls against users, they have maximum independence in deciding how much share will they contribute to the network. CDNs are dedicated networks and so, individual nodes have no autonomy at all. Tightly coupled databases retain all control over the individual sites but however, DDBs created through federating existing databases retain control over local operations.

**Heterogeneity** - Heterogeneity can also be split into many types depending on the differences at various levels of the network stack. *Hardware heterogeneity* is supported by all the wide-area data networks including CDNs [18]. *System heterogeneity* covers heterogeneity at the operating system level and is observed in all the presented data-intensive networks. *Protocol heterogeneity* is supported only by Data Grids as the others require common protocols for communication and data transfer. *Representation heterogeneity* refers to the heterogeneity of the file formats that are supported by the data-intensive networks. Data Grids, P2P Networks CDNs and distributed databases all support exchange of different data formats.

**Management Entity**   - The management entity administers the tasks for maintaining the aggregation. Generally, this entity is a collection of the stakeholders within the distribution network. While this body usually does not have control over individual nodes, nevertheless, it provides services such as a common data directory for locating content and an authentication service for the users of the network. For the Data Grid, we have already discussed the concept of VO. Though entities in a P2P network are independent, a central entity may provide directory service as in the case of Napster. CDNs are owned and maintained by a corporation or a single organisation. Likewise, DDBs are also maintained by single organisations eventhough the consituent databases may be independent.

**Security Requirements**   - Security requirements differ depending on perspective. In a data distribution network, security may have to be ensured against corruption of content (data integrity), for safeguarding users' privacy (anonymity) and for resources to verify users' access rights (authentication). As said before, Data Grids have fine-grained access permissions and require verification of users' identity. P2P Networks such as Freenet are more concerned with preserving anonymity of the users as they may be breaking local censorship laws. A CDN primarily has to verify data integrity since the only access granted is to the content provider. Users have to authenticate against a DDB for carrying out queries and transactions and data integrity has to be maintained for deterministic operation.

**Latency Management & Performance**   - A key element of performance in distributed data-intensive networks is the manner in which they reduce the latency of data transfers. Some of the techniques commonly used in this regard are replicating data close to the point of consumption, caching of data, streaming data, pre-staging the data before the application starts executing and optimally selecting data and computational resources when they are dispersed. In Data Grids, all of the above techniques are implemented in one form or another. CDNs employ caching and streaming to enhance performance especially for delivering media content [52]. Similarly, P2P networks also employ replication, caching and streaming of data in various degrees. Replication and caching are used in distributed database systems for optimizing distributed query processing [53].

Thus, it can be seen that though Data Grids share many characteristics with other types of data intensive network computing technologies, they are differentiated by heavy computational requirements, wider heterogeneity and autonomy and the presence of VOs. Most of the current Data Grid implementations focus on scientific applications. Recent approaches have, however, explored the integration of the above-mentioned technologies within Data Grids to take advantage of the strengths that they offer in areas such as data discovery, storage management and data replication. This is possible as Data Grids already encompass and build on diverse technologies. Foster and Iamnitchi [54] discuss the convergence of P2P and Grid computing and contend that the latter will be able to take advantage of the failure resistance and scalability offered by the former who gains from the experience in managing diverse and powerful resources, complex applications and the multitude of users with different requirements. Ledlie et al. [55] present a similar view and discuss the areas of aggregation, algorithms and maintenance where P2P research can be beneficial to Grids. Practical Grid technologies such as Narada Brokering [56] have used P2P methods for delivering a scalable event-service.

# 3   Taxonomy

This section details a taxonomy that cover various aspects of Data Grids. As Data Grids consist of several elements, our taxonomy covers each one of them in depth. This taxonomy is split into developing four sub-taxonomies as shown in Figure 3. The first sub-taxonomy is from the point of view of Data Grid organization. This classifies all the scientific Data Grids that are currently being developed worldwide. The next sub-taxonomy deals with the transport technologies used within Data Grids. This not only covers well-known file transfer protocols but also includes other means of managing data transportation. A scalable, robust and intelligent replication mechanism is crucial to the smooth operation of a Data Grid and the sub-taxonomy presented next takes into account concerns of Grid environments such as metadata and nature of data transfer mechanisms used. The last sub-taxonomy categorizes resource allocation and scheduling research and looks into issues such as locality of data.

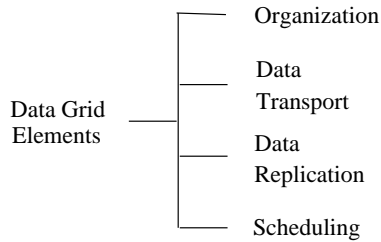

Figure 3: Data Grid Elements.

## 3.1 Data Grid Organization

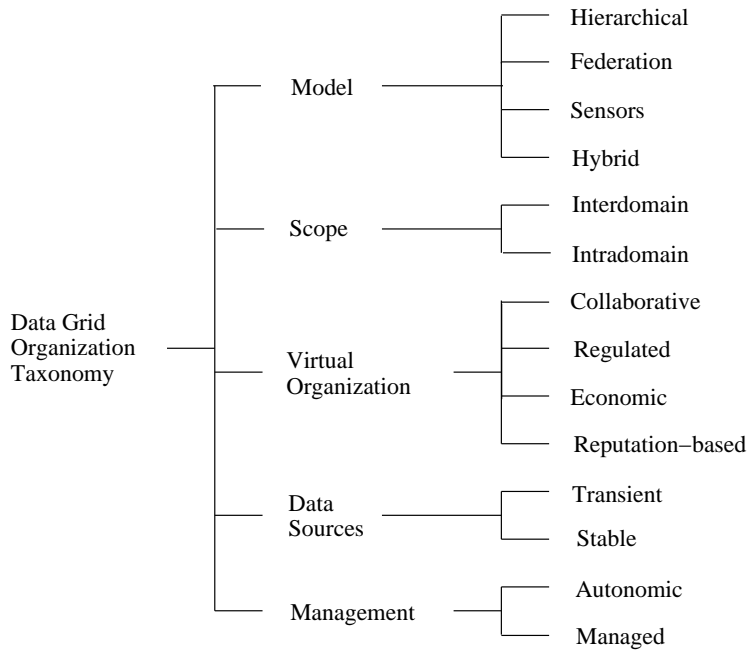

Figure 4: Data Grid Organization Taxonomy.

Figure 4 shows a taxonomy based on the various organizational characteristics of Data Grid projects. These characteristics are central to any Data Grid and are manifest in different ways in different systems.

*Model* - The model is the manner in which data sources are organised in a system. A variety of models are in place for the operation of a Data Grid. These are dependent on: the source of data, whether single or distributed, the size of data and the mode of sharing. Four of the common models found in Data Grids are shown in Figure 5 and are discussed as follows:

- *Hierarchical:* This model is used in Data Grids where there is a single source for data and the data has to be distributed across collaborations worldwide worldwide. For example, the MONARC(Models of Networked Analysis at Regional Centres) group within CERN has proposed a tiered infrastructure model for distribution of CMS data [57]. This model is presented in Figure 5(a) and specifies requirements for transfer of data from CERN to various groups of physicists around the world. The first level is the compute and storage farm at CERN which stores the data generated from the detector. This data is then distributed to sites distributed worldwide called Regional Centres (RCs). From the RCs, the data is then passed downstream to the national and institutional centres and finally onto the physicists. A Tier1 or a Tier2 centre has to satisfy certain bandwidth, storage and computational requirements as shown in the figure.

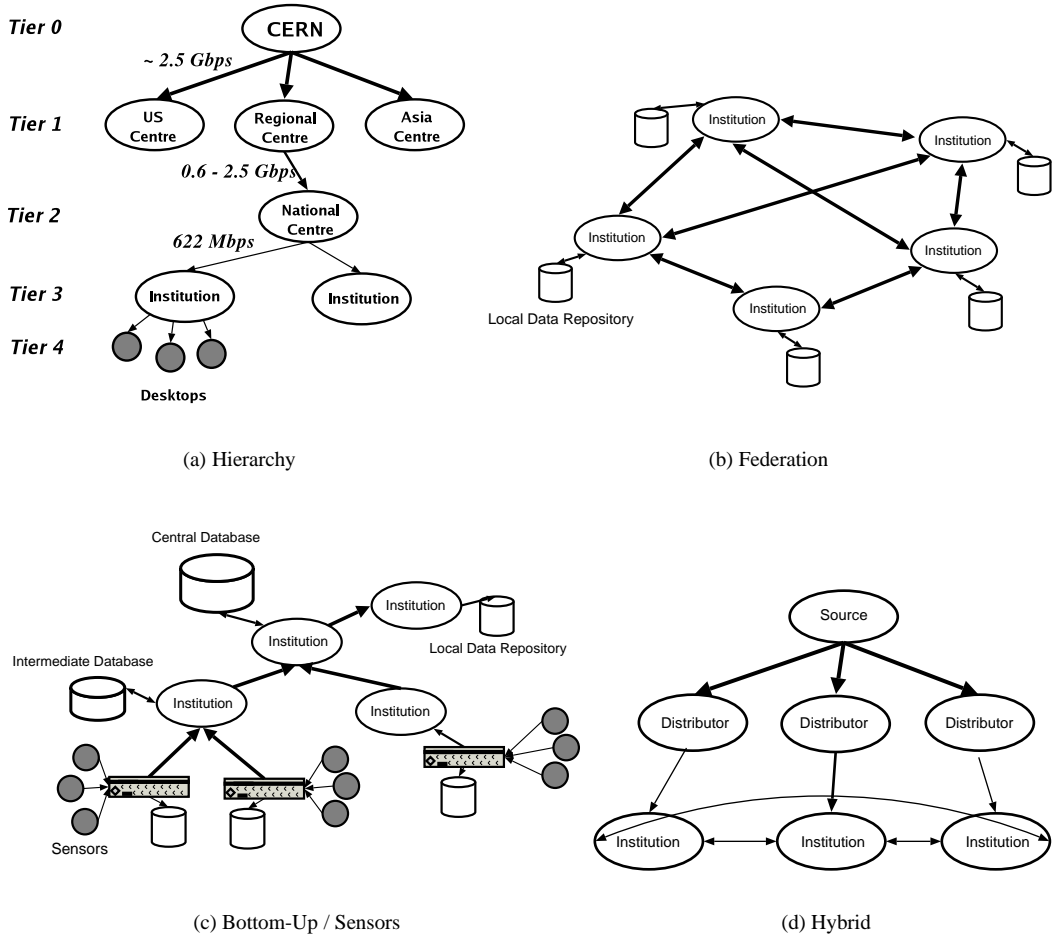

Figure 5: Possible models for organization of Data Grids.

- *Federation:* The federation model [58] is presented in Figure 5(b) and is prevalent in Data Grids created by institutions who wish to share data in already existing databases. One example of a federated Data Grid is the BioInformatics Research Network (BIRN) [45] in United States. Researchers at a participating institution can request data from any one of the databases within the federation as long as they have the proper authentication. Each institution retains control over its local database. Varying degrees of integration can be present within a federated Data Grid. For example, Moore, et. al[59] discuss about 10 different types of federations that are possible using the Storage Resource Broker(SRB) in various configurations. The differences are based on the degree of autonomy of each site, constraints on cross-registration of users, degree of replication of data and degree of synchronization.

- *Bottom-Up / Sensors:* Within this model, the data from sensors distributed worldwide for a particular experiment is gathered at a central database that is queried by researchers. This model is shown in Figure 5(c) and has been applied in the NEESgrid(Network for Earthquake Engineering Simulation) project [60] in the United States. Here the flow of data is from the bottom(sensors) to the top(database). The data is made available through a centralised interface such as a web portal which also verifies users and checks for authorization.

- *Hybrid:* Hybrid models that combine the above models are beginning to emerge as Data Grids mature and enter into production usage. These come out of the need for researchers to collaborate and share products of their analysis. A hybrid model of a hierarchical Data Grid with peer linkages at the edges is shown in Figure 5(d).

*Scope* - The scope of the Data Grid can vary depending on whether it is restricted to a single domain(*intradomain*) or if it is a common infrastructure for various scientific areas(*interdomain*). In the former case, the infrastructure is adapted to the particular needs of that domain. For example, special analysis software may be made available to the participants of a domain-specific Data Grid. In the latter case, the infrastructure provided will be generic.

*Virtual Organizations* - Data Grids are formed by VOs and therefore, the design of VOs reflects on the social organization of the Data Grid. A VO is *collaborative* if it is created by entities who have come together to share resources and collaborate on a single goal. Here, there is an implicit agreement between the participants on the usage of resources. A *regulated* VO may be controlled by a single organization which lays down rules for accessing and sharing resources. In an *economy-based* VO, resource providers enter into collaborations with consumers due to profit motive. In such cases, service-level agreements dictate the rights of each of the participants. A *reputation-based* VO may be created by inviting entities to join a collaboration based on the level of services that they are known to provide.

*Data Sources* - Data sources in a Data Grid may be *transient* or *stable*. A scenario for a transient data source is a satellite which broadcasts data only at certain times of the day. In such cases, applications need to be aware of the short life of the data stream. As we will see later, most of the current Data Grid implementations all have always-on data sources such as mass storage systems or production databases. In future, with diversification, Data Grids are expected to handle transient data sources also.

*Management* - The management of a Data Grid can be *autonomic* or *managed*. Present day Data Grids require plenty of human intervention for tasks such as resource monitoring, user authorization and data replication. However, research is leading to autonomic [61][62] or self-organizing, self-governing Data Grids.

## 3.2   Data Transport

The data transport mechanism is one of the fundamental technologies underlying a Data Grid. Data transport involves not just movement of bits across resources but also other aspects of data access such as security, access controls and management of data transfers. A taxonomy for data transport mechanisms within Data Grids is shown in Figure 6.

*Functions* - Data transport in Grids can be modelled as a three-tier structure that is similar to the networking stacks such as the OSI reference model . At the bottom is the *Transfer Protocol* that specifies a common language for two nodes in a network to initiate and control data transfers. This tier takes care of simple bit movement between two hosts on a network. The most widely-used transport protocols in Data Grids are FTP (File Transfer Protocol) [63] and GridFTP [64]. The second tier is an optional *Overlay Network* that takes care of routing the data. An overlay network provides its own semantics over the Internet protocol to satisfy a particular purpose. In P2P networks, overlays based on distributed hash tables provide a more efficient way of locating and transferring files  [65]. Overlay networks in Data Grids provide services such as storage in the network, caching of data transfers for better reliability and the ability for applications to manage transfer of large datasets. The topmost tier provides application-specific functions such as *File I/O*. A file I/O mechanism allows an application to access remote files as if they are locally available. This mechanism presents to the application a transparent interface through APIs that hide the complexity and the unreliability of the networks. A data transport mechanism can therefore perform one of these functions.

*Security* - Security is an important requirement while accessing or transferring files to ensure proper authentication of users, file integrity and confidentiality. Security can be divided into two main categories: *authentication* of users and *encryption* of data transfer. Authentication can be based on either *passwords* or symmetric or asymmetric *public key* cryptographic protocols such as Kerberos [66] or Grid Security Infrastructure (GSI) [67]. Data encryption may be present or absent within a transfer mechanism. The most prevalent form of data encryption is through SSL (Secure Sockets Layer)[68]. Another aspect of security is the level of *access controls* on the data that is to be transferred. *Coarse-grained* access controls use traditional methods such as UNIX file permissions to restrict the number of files or collections that are accessible to the user. However, expansion of Data Grids to fields such as medical research that have strict controls on the distribution of data have led to requirements for *fine-grained* access controls. Such requirements include usage of tickets for restricting the number of accesses even for authorised users, delegating read and write access rights to particular files or collections and flexible ownership of data [59].

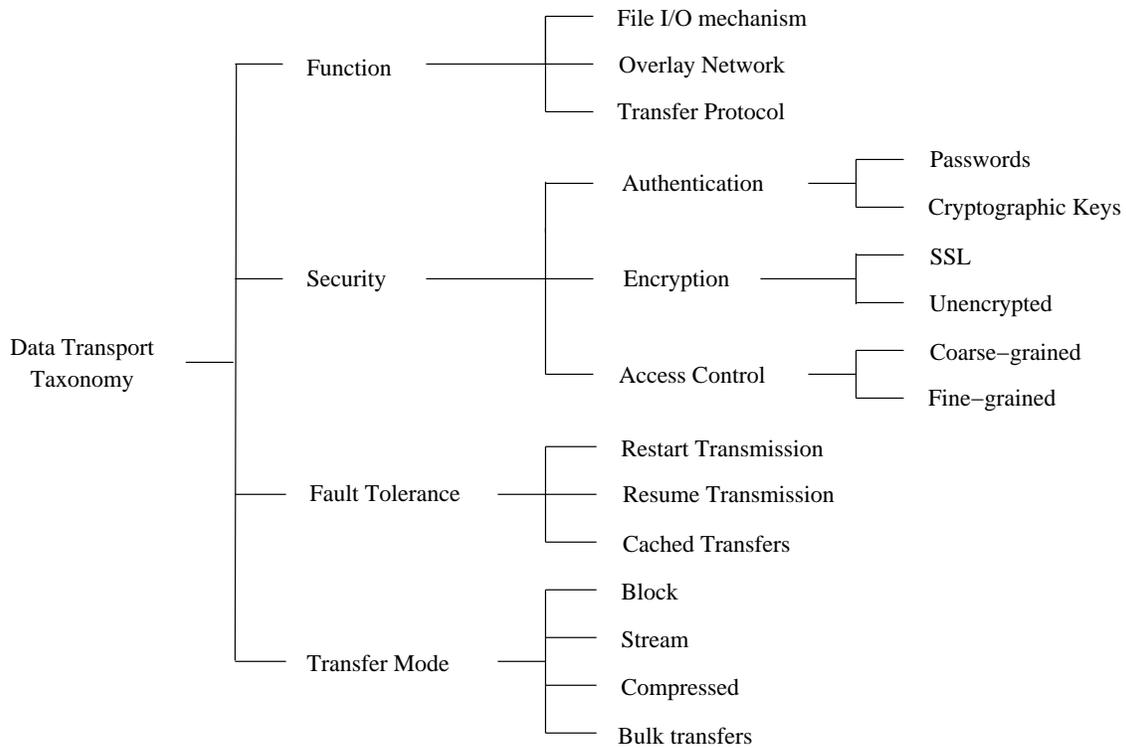

Figure 6: Data Transport Taxonomy.

*Fault Tolerance* - Fault tolerance is also an important feature that is required in a Data Grid environment especially when transfers of large data files occur. Fault tolerance can be subdivided into restarting over, resuming from interruption and providing caching. *Restarting* the transfer all over again means that the data transport mechanism does not provide any failure tolerance. However, all data in transit would be lost and there is a slight overhead for setting up the connection again. Protocols such as GridFTP allow for *resuming* transfers from the last byte acknowledged. Overlay networks provide *caching* of transfers via store-and-forward protocols. In this case, the receiver does not have to wait until the connections are restored. However, caching reduces performance of the overall data transfer and the amount of data that can be cached is dependent on the storage policies at the intermediate network points.

*Transfer Mode* - The last category is the transfer modes supported by the mechanism. The data transfer can be in either *block*, *stream*, *compressed* or *bulk transfer* mode. The last mode is used for latency management while transferring a large amount of data. A mechanism may support more than one mode.

## 3.3 Data Replication and Storage

A Data Grid is a geographically-distributed collaboration in which all members require access to the datasets produced within the collaboration. Replication of the datasets is therefore a key requirement to ensure scalability of the collaboration, reliability of data access and to preserve bandwidth consumption. Replication is bounded by the size of storage available at different sites within the Data Grid and the bandwidth between these sites. A replica management system therefore, ensures access to the required data while managing the underlying storage.

A replica management system, shown in Figure 7, consists of storage nodes which are linked to each other via high-performance data transport protocols. The replica manager directs the creation and management of replicas according to the demands of the users and the availability of storage and a catalog or a directory keeps track of the replicas and their locations. The catalog can be queried by applications to discover the number and the locations of available replicas of a particular dataset. In some systems, the manager and the catalog are merged into one entity. Client-side software generally consists of a library that can be integrated into applications and a set of commands or GUI utilities that are built on top of the libraries. The client libraries allow querying of the catalog to discover datasets and to request replication of a particular dataset.

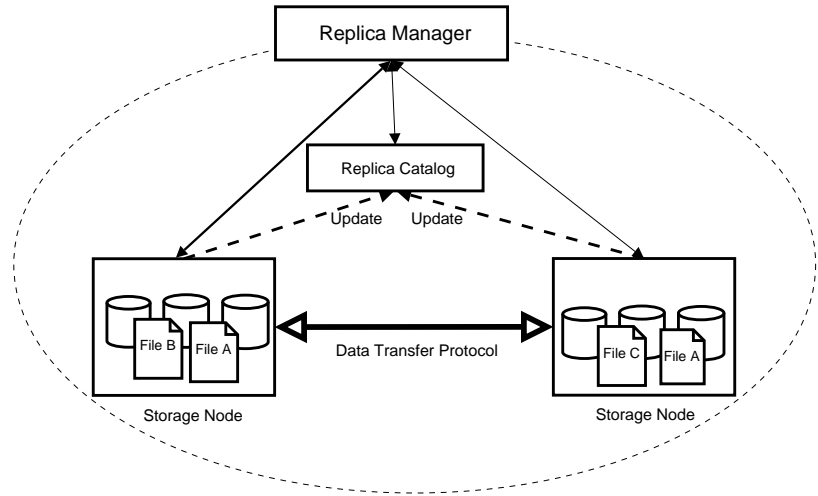

Figure 7: A Replica Management Architecture.

The important elements of a replication mechanism are therefore the architecture of the system and the strategy followed for replication. The first categorization of Data Grid replication is therefore, based on these properties as is shown in Figure 8. The architecture of a replication mechanism can be further subdivided into the categories shown in Figure 9.

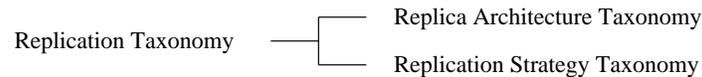

Figure 8: Replication Taxonomy.

*Model & Topology* - The model followed by the system largely determines the way in which the nodes are organized and the method of replication. A *centralized* system would have one master replica which is updated and the updates are propagated to the other nodes. A *decentralized* or peer-to-peer mechanism would have many copies, all of which need to be synchronized with each other. Nodes under a replica management system can be organised either in a ring or a tree topology or in a hybrid of both. The last can be achieved in situations such as a tree with rings at different hierarchies as has been discussed in [69].

*Storage Integration* - The relation of replication to storage is very important and determines the scalability, robustness, adaptability and applicability of the replication mechanism. *Tightly-coupled* replication mechanisms that exert fine-grained control over the replication process are tied to the storage architecture on which they are implemented. The replication system controls the filesystem and I/O mechanism of the local disk. The replication mechanism is therefore at the kernel-level of the OS. Process-based replication is only possible within such a system. Since replication is at the process level, it is often invisible to applications and users. An example of such a mechanism is Gfarm [70]. *Intermediately-coupled* replication systems exert control over the replication mechanism but not over the storage resources. The filesystems are hosted on diverse storage architectures and are controlled by their respective systems. However, the replication is still initiated and managed by the mechanism and therefore, it interacts with the storage system at a very low-level. Such mechanisms work at the application-level and their processes are user-level. Metadata capability is present in this system so that applications or users can request for certain replicas. The data transfer is handled by the system. Therefore, the replication is partially visible. Example of such a system is the Storage Resource Bro-ker (SRB) [71]. *Loosely-coupled* replication mechanisms are superimposed over the existing filesystems and storage systems. The mechanism exerts no control over the filesystem. Replication is initiated and managed by applications and users. Metadata capability is essential to such a system. Such mechanisms interact with the storage systems through standard file transfer protocols and at a high level. The architecture is capable of complete heterogeneity.

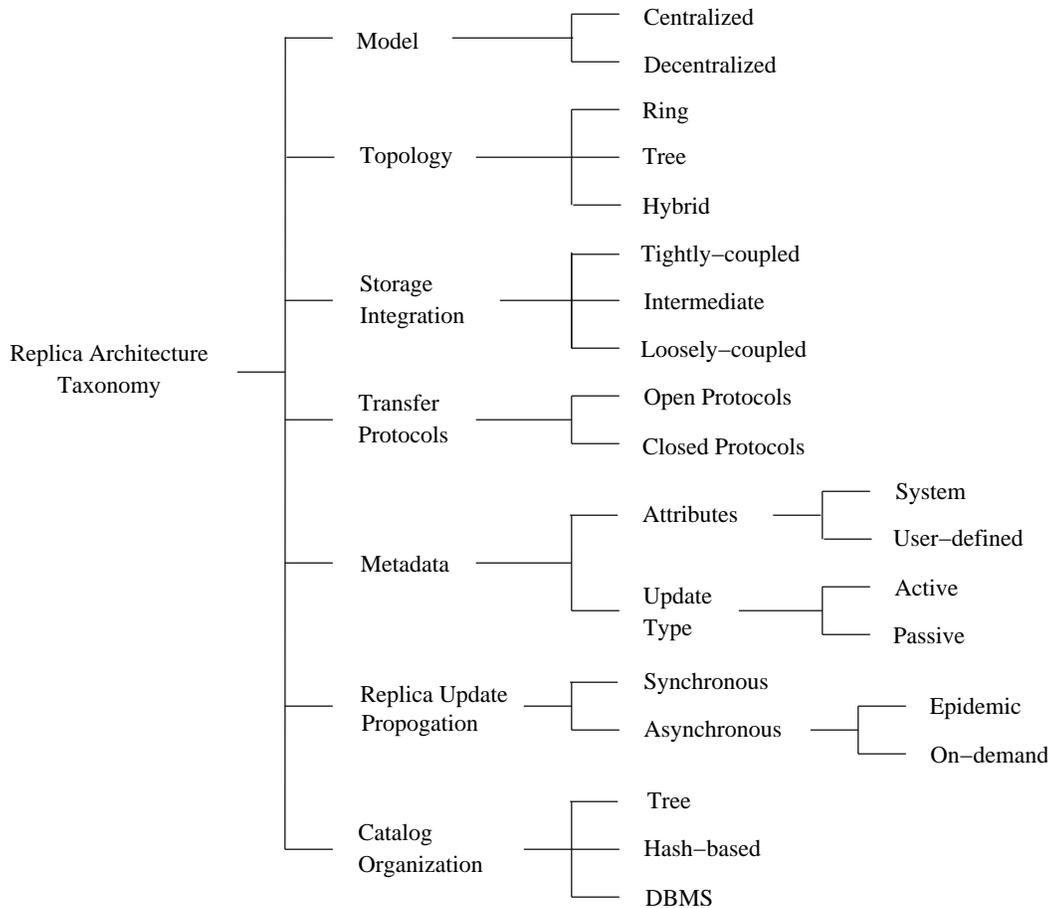

Figure 9: Replica Architecture Taxonomy.

*Transfer Protocols* - The data transport protocols used within replica management systems is also a differentiating characteristic. *Open protocols* for data movement such as GridFTP allow clients to transfer data independent of the replica management system. The replicated data is accessible outside of the replica management system. Systems that follow *closed* or obscure (not widely published) protocols restrict access to the replicas to their client libraries. Tightly-coupled replication systems are mostly closed in terms of data transfer. RLS (Replica Location Service) [72] and GDMP (Grid Data Mirroring Pilot) [73] use GridFTP as their primary transport mechanism. But the flip-side to having open protocols is that the user or the application must take care of updating the replica locations in the catalog if they transfer data outside the replication management system.

*Metadata* - It is difficult, if not impossible, for users to identify particular datasets out of hundreds and thousands that may be present in a large, Grid distributed, collection. From this perspective, having proper metadata or data about the replicated data becomes very important as this allows easy searching based on attributes that are more familiar to the users. At the very least, metadata is mapping of the logical name of the file in the replica manager to the actual physical file on the disk. Metadata can have two types of attributes: one is *system-dependent* metadata, which consists of file attributes such as creation date, size on disk, physical location(s) and file checksum and the other is *user-defined* attributes which consist of properties that depend on the experiment or VO that the user is associated with. For example in a High-Energy Physics experiment, the metadata could describe attributes such as experiment date, mode of production (simulation or experimental) and event type. The metadata can be *actively* updated by the replica management system or else updated *passively* by the users when they create new replicas, modify existing ones or add a new file to the catalog.

*Replica Update Propogation* - Within a Data Grid, data is generally updated at one site and the updates are then propagated to the rest of its replicas. This can be in *synchronous* or in *asynchronous* modes. While synchronous updating is followed in databases, it is not practiced in Data Grids because of the expensive wide-area locking protocols and the frequent movement of massive data required. Asynchronous updating can be epidemic [74], that is, the primary copy is changed and the updates are propagated to all the other replicas or it can be on-demand as in GDMP [75] wherein replica sites subscribe to update notifications at the primary site and decide themselves when to update their copies.

*Catalog Organization* - A replica catalog can be distinguished on the basis of its organization. The catalog can be organized as a *tree* as in the case of LDAP (Lightweight Directory Access Protocol) based catalogs such as the Globus Replica Catalog [76]. The data can be catalogued on the basis of *document hashes* as has been seen in P2P networks. However, SRB and others follow the approach of storing the catalog within a *database*.

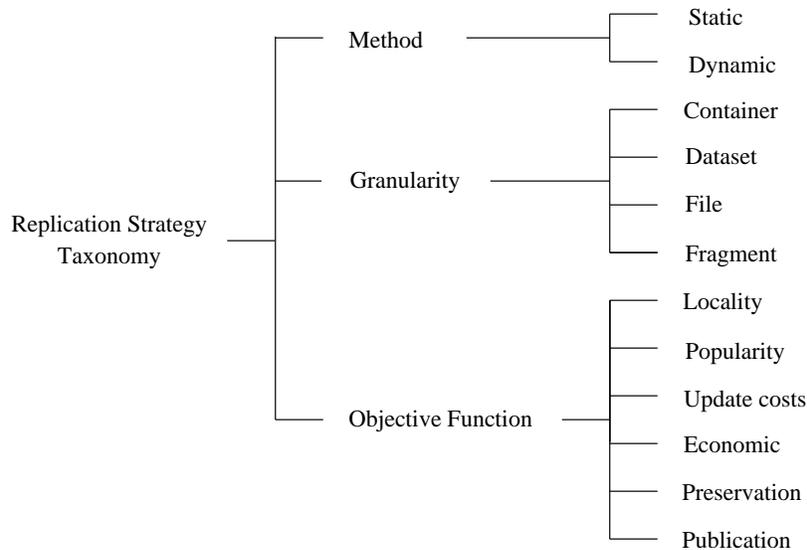

Figure 10: Replication Strategy Taxonomy.

Replication strategies determine when and where to create a replica of the data. These strategies are guided by factors such as demand for data, network conditions and cost of transfer. The replication strategies can be categorized as shown in Figure 10.

*Method* - The first classification is based on whether the strategies are *static* or *dynamic*. Dynamic strategies adapt to changes in demand and bandwidth and storage availability but induce overhead due to more number of operations that they undertake. Also, dynamic strategies are able to recover from failures such as network partitioning. However, frequent transfers of massive datasets that result due to such strategies can lead to strain on the network resources. There may be little gain from using dynamic strategies if the resource conditions are fairly stable in a Data Grid over a long time. Therefore, in such cases, static strategies are applied for replication.

*Granularity* - The granularity of the strategy relates to the level of subdivision of data that the strategy works with. Replication strategies that deal with multiple files at the same time work at the granularity of *datasets*. The next level of granularity is individual *files* while there are some strategies that deal with smaller subdivisions of files such as objects or *fragments*.

*Objective Function* - The third classification deals with the objective function of the replication strategy. Possible objectives of a replication strategy are to maximise the *locality* or move data to the point of computation, to exploit *popularity* by replicating the most requested datasets, to minimize the *update costs* or to maximize some *economic* objective such as profits gained by a particular site for hosting a particular dataset versus the expense of leasing the dataset from some other site. *Preservation* driven strategies provide protection of data even in the case of failures such as corruption or obsolescence of underlying storage media or software errors. Another objective possible for a replication strategy is to ensure effective *publication* by propagating new files to interested clients.

## 3.4   Resource Allocation and Scheduling

The requirements for large datasets and the presence of multiple replicas of these datasets scattered at geographically-distributed locations makes scheduling of data-intensive jobs different from that of computational jobs. Schedulers have to take into account the bandwidth availability and the latency of transfer between a computational node to which a job is going to be submitted and the storage resource(s) from which the data required is to be retrieved. Therefore, the scheduler needs to be aware of any replicas close to the point of computation and if the replication is coupled to the scheduling, then create a new copy of the data.A taxonomy for scheduling of data-intensive applications is shown in Figure 11. The categories are explained as follows:

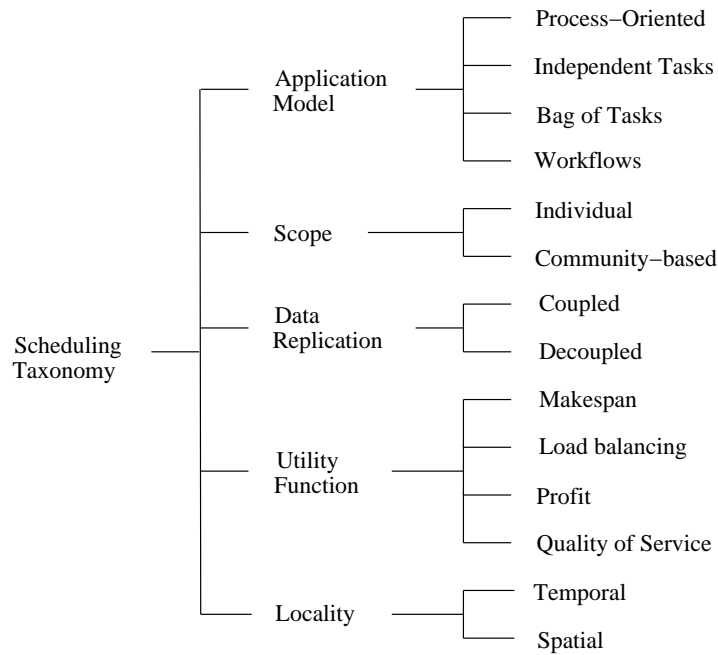

Figure 11: Data Grid Scheduling Taxonomy.

*Application Model* - Scheduling strategies can be classified by the application model that they are targeted towards. Application models are defined in the manner in which the application is composed or distributed for scheduling over global grids. These can range from fine-grained levels such as processes to coarser levels such as individual tasks to sets of tasks such as workflows. Here, a task is considered as the smallest independent unit of computation. Each level has its own scheduling requirements. *Process-oriented* applications are those in which the data is manipulated at the process level. Examples of such applications are MPI programs that execute over global grids. *Independent tasks* having different objectives are scheduled individually and it is ensured that each of them get their required share of resources. A *Bag of Tasks (BoT)* application consists of a set of independent tasks all of which must be executed successfully subject to certain common constraints such as a deadline for the entire application. Such applications arise in parameter studies [77] wherein a set of tasks is created by running the same program on different inputs. In contrast, a *workflow* is a sequence of tasks in which each task is dependent on the results of its predecessor(s). The products of the preceding tasks may be large datasets themselves(for example, a simple two-step workflow could be a data-intensive simulation task and the task for analysis of the results of simulation). Therefore, scheduling of individual tasks in a workflow requires careful analysis of the dependencies and the results to reduce the amount of data transfer.

*Scope* - Scope relates to the extent of application of the scheduling strategy within a Data Grid. If the scope is *individual*, then the scheduling strategy is concerned only with meeting the objectives from a user's perspective. In a multi-user environment therefore, each scheduler would have its own independent view of the resources that it wants to utilise. A scheduler is aware of fluctuations in resource availability caused by other schedulers submitting their jobs to common resources and it strives to schedule jobs on the least-loaded resources that can meet its objectives. With the advent of VOs, efforts have moved towards *community-based* scheduling in which schedulers follow policies that are set at the VO level and enforced at the resource level through service level agreements and allocation quotas [78][79].

*Data Replication* - The next classification relates to whether the job scheduling is *coupled* to data replication or not. Assume a job is scheduled to be executed at a particular compute node. When job scheduling is coupled to replication and the data has to be fetched from remote storage, the scheduler creates a copy of the data at the point of computation so that future requests for the same file that come from the neighbourhood of the compute node can be satisfied more quickly. Not only that, in the future, any job dealing with that particular data will be scheduled at that compute node if available. However, one requirement for a compute node is to have enough storage to store all the copies of data. While storage management schemes such as LRU (Least Recently Used) and FIFO (First In First Out) can be used to manage the copies, the selection of compute nodes is prejudiced by this requirement. There is a possibility that promising computational resources may be disregarded due to lack of storage space. Also, the process of creation of the replica and registering it into a catalog adds further overhead to job execution. In a decoupled scheduler, the job is scheduled to a suitable computational resource and a suitable replica location is identified to request the data required. The storage requirement is transient, that is, disk space is required only for the duration of execution. A comparison of decoupled against coupled strategies in [80] has shown that decoupled strategies promise increased performance and reduce the complexity of designing algorithms for Data Grid environments.

*Utility function* - A job scheduling algorithm tries to minimize or maximize some form of a utility function. The utility function can vary depending on the requirements of the users and architecture of the distributed system that the algorithm is targeted at. Traditionally, scheduling algorithms have aimed at reducing at the total time required for computing all the jobs in a set, also called its *makespan*. *Load balancing* algorithms try to distribute load among the machines so that maximum work can be obtained out of the systems. Scheduling algorithms with economic objectives try to maximize the users' economic utility usually expressed as some *profit* function that takes into account economic costs of executing the jobs on the Data Grid. Another objective possible is to meet the *Quality-of-Service(QoS)* requirements specified by the user. QoS requirements that can be specified include minimising the cost of computation, meeting a deadline, meeting stricter security requirements and/or meeting specific resource requirements.

*Locality* - Exploiting the locality of data has been a tried and tested technique for scheduling and load-balancing in parallel programs [81, 82, 83] and in query processing in databases [84, 85]. Similarly, data grid scheduling algorithms can be categorized as whether they exploit the *spatial* or *temporal* locality of the data requests. Spatial locality is locating a job in such a way that all the data required for the job is available on data hosts that are located close to the point of computation. Temporal locality exploits the fact that if data required for a job is close to a compute node, subsequent jobs which require the same data are scheduled to the same node. Spatial locality can also be termed as "moving computation to data" and temporal locality can be called as "moving data to computation". It can be easily seen that schedulers which couple data replication to job scheduling exploit the temporal locality of data requests.

# 4 Mapping of Taxonomy to Various Data Grid Systems

In this section, we classify various Data Grid research projects according to the taxonomies we developed in Section 3. While the list of example systems is not exhaustive, it is representative of the classes that have been discussed. The projects in each category have been chosen based on several factors such as broad coverage of application areas, project support for one or more applications, scope and visibility, large-scale problem focus and ready availability of documents from project web pages and other sources.

## 4.1 Data Grid Projects

In this space, we study and analyse the various Data Grid projects that have been developed for various application domains around the world. While many of these projects cover aspects of Data Grid research such as middleware development, advanced networking and storage management, we will however, only focus on those projects which are involved in setting up infrastructure for specific areas. A list of these projects and a brief summary about each of them is provided in Table 2. These are also classified according to the taxonomy provided in Figure 4

Table 2: Data Grid Projects around the world.

| Name | Domain | Grid Type | Remarks | Country / Region |
|---|---|---|---|---|
| LCG [86] | High Energy Physics | Hierarchical model, Intradomain, Collaborative VO, Stable Sources, Managed | To create and maintain a data movement and analysis infrastructure for the users of LHC. | Global |
| EGEE [87] | High Energy Physics, Biomedical Sciences | Hierarchical model, Interdomain, Collaborative VO, Stable Sources, Managed | To create a seamless common Grid infrastructure to support scientific research. | Global |
| BIRN [45] | BioInformatics | Federated model, Intradomain, Collaborative VO, Stable Sources, Managed | To foster collaboration in biomedical science through sharing of data. | United States |
| NEESgrid [60] | Earthquake Engineering | Sensor model, Intradomain, Collaborative VO, Transient Sources, Managed | To enable scientists to carry out experiments in distributed locations and analyse data through a uniform interface. | United States |
| GriPhyn [88], PPDG [89] | High Energy Physics | Hierarchical model, Intradomain, Collaborative VO, Stable Sources, Managed | To create an integrated infrastructure that provides computational and storage facilities for high-energy physics experiments. | United States |
| Grid3 [90] | Physics, Biology | Hierarchical model, Interdomain, Collaborative VO, Stable Sources, Managed | To provide a uniform, scalable and managed grid infrastructure for science applications | United States |
| BioGrid, Japan [91] | Protein Simulation, Brain Activity Analysis | Federated model, Intradomain, Collaborative VO, Stable Sources, Managed | Computational and data infrastructure for medical and biological research. | Japan |
| Virtual Observatories [92, 93] | Astronomy | Federated model, Intradomain, Collaborative VO, Stable Sources, Managed | Infrastructure for accessing diverse astronomy observation and simulation archives through integrated mechanisms. | Global |
| Earth System Grid [94] | Climate Modelling | Federated model, Intradomain, Collaborative VO, Stable Sources, Managed | Integrating computational, data and analysis resources to create environment for next generation climate research. | United States |

| GridPP [95] | High Energy Physics | Hierarchical model, Intradomain, Collaborative VO, Stable Sources, Managed | To create computational and storage infrastructure for Particle Physics in the UK. | United Kingdom |
|---|---|---|---|---|
| eDiaMoND [44] | Breast Cancer Treatment | Federated model, Intradomain, Collaborative VO, Stable Sources, Managed | To provide medical professionals and researchers access to distributed databases of mammogram images. | United Kingdom |
| Belle Analysis Data Grid [96] | High Energy Physics | Hierarchical model, Intradomain, Collaborative VO, Stable Sources, Managed | To create computational and storage infrastructure in Australia for physicists involved in the Belle and ATLAS experiments. | Australia |

Some of the scientific domains that are making use of Data Grids are as follows:

**High Energy Physics(HEP)** - The computational and storage requirements for HEP experiments have already been covered in previous literature [7]. Other than the four experiments at the LHC already mentioned, the Belle experiment at KEK, Japan, the BaBar experiment at the Stanford Linear Accelerator Center (SLAC) and the CDF and D0 experiments at Fermi National Laboratory, US are also adopting Data Grid technologies for their computing infrastructure. There have been numerous Grid projects around the world that are setting up the infrastructure for physicists to process data from HEP experiments. Some of these are the LHC Computing Grid (LCG) led by CERN, the Particle Physics Data Grid (PPDG) and Grid Physics Network(GriPhyN) in the United States, GridPP in the UK and Belle Analysis Data Grid (BADG) in Australia. These projects have common features such as a tiered model for distributing the data, shared facilities for computing and storage and personnel dedicated towards managing the infrastructure. Some of them are entering or are being tested for production usage.

**Astronomy** - The community of astrophysicists around the globe are setting up Virtual Observatories for accessing the data archives that has gathered by telescopes and instruments around the world. These include the National Virtual Observatory(NVO) in the US, Australian Virtual Observatory, Astrophysical Virtual Observatory in Europe and AstroGrid in the UK. The International Virtual Observatory Alliance (IVOA) is coordinating these efforts around the world for ensuring interoperability. Commonly, these projects provide uniform access to data repositories along with access to software libraries and tools that may be required to analyse the data. Other services that are provided include access to high-performance computing facilities and visualization through desktop tools such as web browsers. Other astronomy grid projects include those being constructed for the LIGO (Laser Interferometer Gravitational-wave Observatory) [97] and SDSS (Sloan Digital Sky Survey) [98] projects.

**BioInformatics** - The increasing importance of realistic modeling and simulation of biological processes coupled with the need for accessing existing databases has led to Data Grid solutions being adopted by bioinformatics researchers worldwide. These projects involve federating existing databases and providing common data formats for the information exchange. Examples of these projects are BioGrid project in Japan for online brain activity analysis and protein folding simulation, the eDiaMoND project in UK for breast cancer treatment and the BioInformatics Research Network (BIRN) for imaging of neurological disorders using data from federated databases.

**Earth Sciences** - Researchers in disciplines such as earthquake engineering and climate modeling and simulation are adopting Grids to solve their computational and data requirements. NEESgrid is a project to link earthquake researchers with high performance computing and sensor equipment so that they can collaborate on designing and performing experiments. Earth Systems Grid aims to integrate high-performance computational and data resources to study the petabytes of data resulting from climate modelling and simulation.

## 4.2    Data Transport Technologies

Within this subsection, various projects involved in data transport over Grids are discussed and classified according to the taxonomy provided in Section 3.2. The data transport technologies studied here range from protocols such as FTP to overlay methods such as Internet Backplane Protocol to file I/O mechanisms. Each of the technologies have unique properties and are representative of the categories in which they are placed. A summary of these technologies and their categorization is provided in Table 3.

Table 3: Comparison between various data transport technologies.

| Project | Function | Security | Fault Tolerance | Transfer Mode |
|---------|----------|----------|-----------------|---------------|
| GASS [99] | File I/O | PKI, Unencrypted, Coarse-grained | Caching | Block, Stream append |
| IBP [100][101] | Overlay Mechanism | Password, Unencrypted, Coarse-grained | Caching | Block |
| FTP [63] | Transfer Protocol | Password, Unencrypted, Coarse-grained | Restart | All |
| SFTP [102] | Transfer Protocol | PKI, SSL, Coarse-grained | Restart | All |
| GridFTP [64][103] | Transfer Protocol | PKI, SSL, Coarse-grained | Resume | All |
| Kangaroo [104] | Overlay Mechanism | PKI, Unencrypted, Coarse-grained | Caching | Block |
| Legion [105] | File I/O | PKI, Unencrypted, Coarse-grained | Caching | Block |
| SRB [71] | File I/O | PKI, SSL, Fine-grained | Restart | Block, Bulk transfer |
| Stork [106] | File I/O | PKI, SSL, Coarse-grained | Resume | Block, Stream |

### 4.2.1    GASS

Global Access to Secondary Storage(GASS) [99] is a data access mechanism provided within the Globus toolkit for reading local data at remote machines and for writing data to remote storage and moving it to a local disk. The goal of GASS is to provide a uniform data access interface to applications running at remote resources while keeping the functionality demands on both the resources and the applications limited. GASS is not an attempt to provide a Grid wide distributed file system but instead is a remote I/O mechanism for Grid applications.

GASS conducts its operations via a file cache which is an area on the secondary storage where the remote files are stored. When a remote file is requested by an application for reading, GASS by default fetches the entire file into the cache from where it is opened for reading as in a conventional file access. It is retained in the cache as long as applications are accessing it. While writing to a remote file, the file is created or opened within the cache where GASS keeps track of all the applications writing to it via reference count. When the reference count is zero, the file is transferred to the remote machine. Therefore, all operations on remote file are conducted locally in the cache which reduces demand on bandwidth. A large file can be *prestaged* into the cache, that is, fetched before an application requests it for reading. Similarly, a file can be transferred out via *poststaging*. It is also possible to transfer files in and out of disk areas other than the caches. GASS operations are available through an API and also through Globus commands. GASS is integrated with the Globus Resource Access and Monitoring(GRAM) service [107] and is used for staging executables, staging in files and retreving the standard output and error streams of the jobs.

GASS provides a limited ability for data transfer between remote nodes. As it prefetches the entire file into the cache, it is not suitable as a transfer mechanism for large data files (of GigaByte upwards) as the required cache capacity might not be available. Also, it does not provide features such as file striping, third-party transfer, TCP tuning, etc. provided by protocols such as GridFTP. However, because of its lightweight functionality, it is suitable for applications where the overhead of setting up a GridFTP connection dominates.

### 4.2.2 IBP

Internet Backplane Protocol (IBP) [100][101] allows applications to optimize data transfer and storage operations by controlling data transfer explicitly by storing it at intermediate locations. IBP uses a "store-and-forward" protocol to move data around the network. Each of the IBP nodes has a temporary buffer into which data can be stored for a fixed amount of time. Applications can manipulate these buffers so that data is moved to locations close to where it is required.

IBP is modelled after the Internet Protocol. The data is handled in units of fixed-size byte arrays which are analogous to IP datagrams or network packets. Just as IP datagrams are independent of the data link layer, byte arrays are independent of the underlying storage nodes. This means that applications can move data around without worrying about managing storage on the individual nodes. IBP also provides a global addressing space that is based on global IP addressing. Thus, any client within an IBP network can make use of any IBP node.

IBP can also be thought of as a virtualisation layer built on top of storage resources (or access layer). IBP provides access to heterogeneous storage resources through a global addressing space in terms of fixed block sizes thus making access to data independent of the storage method and media. The storage buffers can grow to any size and thus, the byte arrays can also be thought of as files which live on the network.

IBP also provides client API and libraries that provide semantics similar to UNIX system calls. A client connects to an IBP "depot", or a server, and requests storage allocation. In return, the server provides it three *capabilities*: for reading from, writing to and managing the allocation. Capabilities are cryptographically secure byte strings which are generated by the server. Subsequent calls from the client must make use of the same capabilities to perform the operations. Thus, capabilities provide a notion of security as a client can only manipulate its own data. Capabilities can be exchanged between clients as they are text. Higher-order aggregation of byte arrays is possible through exNodes which are similar to UNIX inodes. exNodes allow uploading, replicating and managing of files on a network with an IBP layer above the networking layer [108].

However, IBP is a low-level storage solution that functions just above the networking layer. Beyond the use of capabilities, it does not have a address mechanism that keeps track of every replica generated. There is no directory service that keeps track of every replica and no information service that can return the IBP address of a replica once queried. Though exNodes store metadata, IBP itself does not provide a metadata searching service.

### 4.2.3 FTP

FTP (File Transfer Protocol) [63] is one of the fundamental protocols for data movement in the Internet. FTP is therefore ubiquitous and every operating system ships with an FTP client.

FTP separates the process of data transfer into two channels, the control channel used for sending commands and replies between a client and a server and the data channel through which the actual transfer takes place. The FTP commands set up the data connection by specifying the parameters such as data port, mode of transfer, data representation and structure. Once the connection is set up the server then initiates the data transfer between itself and the client. The separation of control and data channels also allows third-party transfers to take place. A client can open two control channels to two servers and direct them to start a data transfer between themselves bypassing the client. Data can be transferred in three modes : stream, block and compressed. In the stream mode, data is transmitted as is and it is the responsibility of the sending host to notify the end of stream. In the block mode, data is transferred as a series of blocks preceded by header bytes. In the compressed mode, a preceding byte denotes the number of replications of the following byte and filler bytes are represented by a single byte.

Error recovery and restart within FTP does not cover corrupted data but takes care of data lost due to loss of network or a host or of the FTP process itself. This requires the sending host to insert markers at regular intervals within the data stream. A transmission is restarted from the last marker sent by the sender before the previous transfer crashed. However, restart is not available within the stream transfer mode. Security within FTP is very minimal and limited to the control channel. The username and password are transmitted as clear text and there is no facility for encrypting data while in transit within the protocol. This limits the use of FTP for confidential transfers.

Numerous extensions to FTP have been proposed to offset its limitations. RFCs 2228 [109] and 2389 [110] propose security and features extensions to FTP respectively. However, these are not implemented by popular FTP servers such as wu-ftpd. SSH File Transfer Protocol (SFTP) [102] is a secure file transfer protocol that uses the Secure Shell (SSH) Protocol for both authentication and data channel encryption. SFTP is designed to be both a transfer protocol and a remote file system access protocol. However, it does not support features required for high-performance data transfer such as parallel and striped data transfer, resuming interrupted transmissions or tuning of TCP windows.

### 4.2.4 GridFTP

GridFTP [64][103] extends the default FTP protocol by providing features that are required in a Data Grid environment. The aim of GridFTP is to provide secure, efficient, and reliable data transfer in Grid environments.

GridFTP extends the FTP protocol by allowing GSI and Kerberos based authentication. GridFTP provides mechanisms for parallel data transfer, i.e., ability to maximise usage of bandwidth by transferring a file using multiple TCP streams over the same channel between a source and a destination. It allows third-party transfer, i.e., the ability for one site to start and control file transfers between two other sites. Data can be *striped* over different storage nodes, that is, divided up and spread among the disks. GridFTP allows access to these different blocks simultaneously. Striped-data transfers can further improve bandwidth utilisation and speed-up file transfer. GridFTP supports partial file transfer, ability to access only part of a file. It allows changing the sizes of the TCP buffers and congestion windows to improve transfer performance. Transfer of massive data-sets is prone to failures as the network may exhibit transient behaviour over long periods of time. GridFTP sends restart markers indicating a byte range that has been successfully written by the receiver every 5 seconds over the control channel. In case of a failure, transmission is resumed from the point indicated by the last restart marker received by the sender.

GridFTP provides these features by extending the basic FTP protocol through new commands, features and a new transfer mode. The Striped Passive(SPAS) command is an extension to the FTP PASV command wherein the server presents a list of ports to connect to rather than just a single port. This allows for multiple connections to download the same file or for receiving multiple files in parallel. The Extended Retrieve (ERET) command supports partial file transfer among other things. The Set Buffer (SBUF) and AutoNegotiate Buffer (ABUF) extensions allow the resizing of TCP buffers on both client and server sides. The Data Channel Authentication (DCAU) extension provides for encrypting of data channels for confidential file transfer. DCAU is used only when the control channel is authenticated through RFC 2228 mechanisms. Parallel and striped data transfers are realised through a new transfer more called the extended block mode (mode E). The sender notifies the receiver of the number of data streams by using the End of Data (EOD) and End of Data Count (EODC) codes. The EODC code signifies how many EOD codes should be received to consider a transfer closed. An additional protocol is therefore required from the sender side to ensure that the receiver obtains the data correctly. GridFTP implements RFC 2389 for negotiation of feature sets between the client and the server. Therefore, the sender first requests the features supported by the reciever and then sets connection parameters accordingly. GridFTP also supports restart for stream mode transfers which is not provided in the vanilla FTP protocol.

The only public implementation for the GridFTP server-side protocols is provided in the Globus Toolkit [111]. The Globus GridFTP server is a modified wu-ftpd server that supports most of GridFTP features except for striped data transfer and automatic TCP buffer size negotiation. The Globus Toolkit provides libraries and APIs for clients to connect to GridFTP servers. A command-line tool, *globus-url-copy*, built using these libraries functions as a GridFTP client. Other examples of GridFTP clients include the file transfer component [112] within the Java CoG kit [113] and the UberFTP [114] client from NCSA.

Evaluation of GridFTP protocols alongside FTP has shown that using the additional features of GridFTP increases performance of data transfer [115]. Particularly, the usage of parallel threads dramatically improves the transfer speed over both loaded and unloaded networks. Also, parallel transfers saturate the bandwidth thus improving the link utilisation [76].

### 4.2.5 Kangaroo

Kangaroo [104] is an end-to-end data movement protocol that aims to improve the responsiveness and reliability of large data transfers within the Grid. The main idea in Kangaroo is to conduct the data transfer as a background process so that failures due to server crashes and network partitions are handled transparently by the process instead of the application having to deal with them.

Kangaroo uses memory and disk storage as buffers to which data is written to by the application and moved out by a background process. The transfer of data is performed concurrently with CPU bursts thereby improving utilization. The transfer is conducted through *hops*, or stages where an intermediate server is introduced between the client and the remote storage from which the data is to be read or written. Data received by the intermediate stage is spooled into the disk from where it is copied to the next stage by a background process called the *mover*. This means that a client application writing data to a remote storage is isolated from the effects of a network crash or slow-down as long as it can keep writing to the disk spool. However, it is also possible for a client to write data to the destination server directly over a TCP connection using the Kangaroo primitives.

Kangaroo services are provided through an interface which implements four simple file semantics: `get`(non-blocking read), `put`(non-blocking write), `commit`(block until writes have been delivered to the next stage) and `push`(block until all writes are delivered to the final destination). However, this provides only weak consistency since it is envisioned for grid applications in which data flow is primarily in one direction. As can be seen, Kangaroo is an output-oriented protocol which primarily deals with reliability of data transfer between a client and a server.

The design of Kangaroo is similar to that of IBP even though their aims are different. Both of them use store-and-forward method as a means of transporting data. However, while IBP allows applications to explicitly control data movement through a network, Kangaroo aims to keep the data transfer hidden through the usage of background processes. Also, IBP uses byte arrays whereas Kangaroo uses the default TCP/IP datagrams for data transmission.

### 4.2.6 Legion I/O model

Legion [105] is a object-oriented grid middleware for providing a single system image across a collection of distributed resources. The I/O mechanism within Legion [116] aims to provide transparent access to files stored on distributed resources through APIs and daemons that can be used by native and legacy applications alike.

Resources within the Legion system are represented by objects. BasicFileObjects correspond to files in a conventional file system while ContextObjects correspond to directories. However, these are separated from the actual file system. A datafile is copied to a BasicFileObject to be registered within the context space of Legion. The context space provides location-independent identifiers which are bound to human-readable context names. This presents a single address space and hierarchy from which users can request files without worrying about their location. Also, the representation of BasicFileObject is system-independent and therefore, provides interoperability between heterogeneous systems.

Access to a Legion file object is provided through various means. Command-line utilities provide a familiar interface to the Legion context space. Application developers can use APIs which closely mimic C and C++ file primitives and Unix system calls. For legacy codes, a buffering interface is provided through which applications can operate on local files copied from the Legion objects and the changes are copied back. Another method is to use a modified NFS daemon that translates client request to appropriate Legion invocations.

Security for file transfer is provided through means of X.509 proxies which are delegated to the file access mechanisms [117]. Data itself is not encrypted while in transit. Caching and prefetching is implemented for increasing performance and to ensure reliability.

#### 4.2.7 SRB I/O

The Storage Resource Broker(SRB) [71] developed at the San Diego Supercomputing Centre (SDSC) focuses on providing a uniform and transparent interface to heterogeneous storage systems that include disks, tape archives and databases. A study of SRB as a replication mechanism is provided in the following section, in this section however, we will focus on the data transport mechanism within SRB.

Data transport within SRB provides sophisticated features for performing bulk data transfer operations across geographically distributed sites. Two such feature are support for parallel-I/O and third-party transfers. SRB provides for strong security mechanisms supported by fine-grained access controls on data. Access security is provided through credentials such as passwords or public key and private key pair which can be stored within MCAT itself. Controlled authorization for read access is provided through tickets issued by users who have control privileges on data. Tickets are time-limited or use-limited. Users can also control access privileges along a collection hierarchy.

SRB also provides support for remote procedures. These are operations which can be performed on the data within SRB without having to move it. Remote procedures include execution of SQL queries, filtering of data and metadata extraction. This also provides for an additional level of access control as users can specify certain dataets or collections to be accessible only through remote procedures.

#### 4.2.8 Stork

Stork [106] is a scheduler for data placement jobs on the Grid. That is, Stork exclusively manages operations such as locating, accessing, storing and replicating data equivalent to managing computational jobs by introducing services such as queueing and checkpointing of transfers and managing storage requirements. The aim of Stork is to handle data transfers among heterogeneous systems taking care of job priorities and transfer failures while avoiding overloading of network resources.

Stork is implemented over the existing storage systems and data transfer protocols. It can, therefore, negotiate with different kinds of storage systems and middleware and can translate between mutually incompatible file transfer protocols. It can also automatically decide which protocol to use to transfer data from one host to another. Stork uses ClassAds [118] as the mechanism to represent job and data requirements.

If the data cannot be transferred directly, Stork is capable of determining intermediate steps required to complete the transfer and then combining all the intermediate steps to form a DAG(Directed Acyclic Graph) which is then used as a plan for execution to be managed by a higher-level planner. Stork uses intermediate caches to form *data pipelines* [119] in order to transfer data over an indirect path between two hosts.

While Stork provides almost all the functionality of Kangaroo and some of IBP and SRB, the scheduling of data placement jobs makes this system unique. This helps higher-level scheduling of network requirements in order to make full use of the link capability and avoid overloading.

### 4.3 Data Replication and Storage

In this subsection, four of the data replication mechanisms used within Data Grids are studied in depth and classified according to the taxonomy given in Section 3.3. These were chosen not only because of their wide usage but also because of the wide variations in design and implementation that these represent. A summary is given in Table 4. Table 5 encapsulates the differences between the various replication mechanisms on the basis of the replication strategies that they follow. Some of the replication strategies have been only simulated and therefore, these are explained in a separate subsection.

#### 4.3.1 Grid DataFarm

Grid Datafarm(Gfarm) [70] is an architecture that couples storage, I/O bandwidth and processing to provide scalable computing to process petabytes(PB) of data. The architecture consists of nodes that have a large disk space (in the order of terabytes(TB)) coupled with computing power. These nodes are connected via a high speed interconnect such as Myrinet or Fast Ethernet. Gfarm consists of the Gfarm filesystem, process scheduler and the parallel I/O APIs.

Gfarm filesystem is a parallel filesystem that unifies the file addressing space over all the nodes. It provides scalable I/O bandwidth by integrating process scheduling with data distribution. A Gfarm file is a large file that is stored throughout the filesystem on multiple disks as fragments. Each fragment has arbitrary length and can be stored on any node. Individual fragments can be replicated and the replicas are managed through Gfarm metadata. Individual fragments may be replicated and the replicas are managed through the filesystem metadata and replica catalog. Metadata is updated at the end of each operation on a file. A Gfarm file is write-once, that is, if a file is modified and saved, then internally it is versioned and a new file is created.

Table 4: Comparison between various data replication mechanisms.

| Project | Model | Topology | Storage Integration | Data Transport | Metadata | Update Propagation | Catalog Organization |
|---------|-------|----------|--------------------|-----------------|----------|--------------------|--------------------|
| Grid Datafarm [70] | Centralised | Tree | Tightly-coupled | Closed | System, Active | Asynchronous, epidemic | DBMS |
| RLS [72] | Centralised | Tree | Loosely-coupled | Open | User-defined, Passive | Asynchronous, on-demand | DBMS |
| GDMP [73] | Centralised | Tree | Loosely-coupled | Open | User-defined, Passive | Asynchronous, on-demand | DBMS |
| SRB [71] | Decentralised | Hybrid | Intermediate | Closed | User-defined, Passive | Asynchronous, on-demand | DBMS |

Table 5: Comparison between replication strategies.

| Project | Method | Granularity | Objective Function |
|---------|--------|-------------|--------------------|
| Grid Datafarm | Static | File, Fragment | Locality |
| RLS | Static | Datasets, File | Popularity, Publication |
| GDMP [75] | Static | Datasets, File, Fragment | Popularity, Publication |
| SRB | Static | Containers, Datasets, File | Preservation, Publication |
| Lamehamedi, et. al [69, 120] | Dynamic | File | Update Costs |
| Bell, et. al [121] | Dynamic | File | Economic |
| Lee and Weissman [122] | Dynamic | File | Popularity |
| Ranganathan, et. al [123] | Dynamic | File | Popularity |

Gfarm targets data-intensive applications in which the same program is executed over different data files and where the primary task is of reading a large body of data. The size of the data is in the range of terabytes(TB) and therefore, it is split up and stored as fragments on the nodes. While executing a program, the process scheduler dispatches it to the node that has the segment of data that the program wants to access. If the nodes that contain the data and its replicas are under heavy CPU load, then the filesystem creates a replica of the requested fragment on another node and assigns the process to it. In this way, I/O bandwidth is gained by exploiting the access locality of data. This process can also be controlled through the Gfarm APIs. It is also possible to access the file using a local buffer cache instead of replication.

On the whole, Gfarm is a system that is tuned for high-speed data access within a tightly-coupled yet large-scale architecture such as clusters consisting of hundreds of nodes. It requires high-speed interconnects between the nodes so that bandwidth-intensive tasks such as replication do not cause performance hits. This is evident through experiments carried out over clusters and wide-area testbeds [124][125]. The scheduling in Gfarm is at the process level and applications have to use the API though a system call trapping library is provided for inter-operating with legacy applications. Gfarm targets applications such as High Energy Physics where the data is "write-once read-many". For applications where the data is constantly updated, there could be problems with managing the consistency of the replicas and the metadata though an upcoming version aims to fix them [126].

### 4.3.2 RLS

Giggle (GIGa-scale Global Location Engine) [72] is an architectural framework for a Replica Location Service(RLS) that maintains information about physical locations of copies of data. The main components of RLS are the Local Replica Catalog (LRC) which maps the logical representation to the physical locations and the Replica Location Index (RLI) which indexes the catalog itself.

The actual data is represented by a *logical file name (LFN)* and contain some information such as the size of the file, its creation date and any other such metadata that might help users to identify the files that they seek. A logical file has a mapping to the actual physical location(s) of the data file and its replicas, if any. The physical location is identified by a unique *physical file name (PFN)* which is a URL to the data file on storage. Therefore, a LRC provides the PFN corresponding to an LFN. The LRC also supports authenticated queries that is, information about the data is not available in the absence of proper credentials.

A data file may be replicated across several geographical and administrative boundaries and information about its replicas may be present in several replica catalogs. An RLI creates an index of replica catalogs as a set of logical file names and a pointer to a replica catalog entries.Therefore, it is possible to define several configurations of replica indexes, for example a hierarchical configuration or a central, single-indexed configuration or a partitioned index configuration. Some of the possible configurations are listed in [72]. The information within an RLI is periodically updated using soft-state mechanisms similar to those used in Globus MDS. Infact, the replica catalog structure is quite similar to that of GRIS[127].

RLS is aimed at replicating data that is "write once read many". Data from scientific instruments that needs to be distributed around the world is falls into this category. This data is seldom updated and therefore, strict consistency management is not required. Soft-state management is enough for such applications. RLS is also a standalone replication service that is it does not handle file transfer or data replication itself. It provides only an index for the replicated data.

### 4.3.3 GDMP

GDMP [73][75] is a replication manager that aims to provide secure and high-speed file transfer services for replicating large data files and object databases. GDMP leverages the capabilities of other DataGrid tools such as replica catalogs and GridFTP to provide point-point replication capabilities.

GDMP is based on the publish-subscribe model, wherein the server publishes the set of new files that are added to the replica catalog and the client can request for a copy of these after making a secure connection to the server. GDMP uses the Grid Security Infrastructure(GSI) as its authentication and authorization infrastructure. Clients first register with the server and receive notifications about new data that are available which are then requested for replication. Failure during replication is assumed to be handled by the client. For example, if the connection fails while replicating a set of files, the client may reconnect with the server and request for a re-transfer. The file transfer is done through GridFTP.

GDMP deals with object databases created by High Energy Physics experiments. A single file may contain a upto a billion ($10^9$) objects and therefore, it is advantageous for the replication mechanisms to deal with objects rather than files. Objects requested by a site are copied to a new file at the source. This file is then transferred to the recipient and the database at the remote end is updated to include the new objects. The file is then deleted at the origin. In this case, replication is static as changing Grid conditions are not taken into account by the source site. It is left upto the client site to determine when to replicate and which files to replicate.

GDMP was originally conceived for the Compact Muon Solenoid (CMS) experiment at the LHC in which the data is generated at one point and has to be replicated globally. Therefore, consistency of replicas is not a big issue as there are no updates and all the notifications are mostly in a single direction. The data for this experiment was in the form of files containing objects where each object represented a collision. GDMP can interact with the object database to replicate specific groups of objects between sites.

### 4.3.4 Storage Resource Broker (SRB)

The purpose of the SRB is to enable the creation of shared collections through management of consistent state information, latency management, load leveling, logical resources usage and multiple access interfaces [71][128]. SRB also aims to provide a unified view of the data files stored in disparate media and locations by providing the capability to organise them into virtual collections independent of their physical location and organization. It provides a large number of capabilities that are not only applicable to Data Grids but also for collection building, digital libraries and persistent archival applications.

An SRB installation follows a three-tier architecture - the bottom tier is the actual storage resource, the middleware lies in between and at the top is the Application Programming Interface (API) and the metadata catalog (MCAT). File systems and databases are managed as *physical storage resources (PSRs)* which are then combined into *logical storage resources (LSRs)*. Data items in SRB are organised within a hierarchy of collections and sub-collections that is analogous to the UNIX filesystem hierarchy. Collections are implemented using LSRs while the data items within a collection can be located on any PSR. Data items within SRB collections are associated with metadata which describe system attributes such as access information and size and descriptive attributes which record properties deemed important by the users. The metadata is stored within MCAT which also records attributes of the collections and the PSRs. Attribute-based access to the data items is made possible by searching MCAT.

The middleware is made up of SRB Master daemon and the SRB Agent processes. The clients authenticate to the SRB Master and the latter starts an Agent process that processes the client requests. An SRB agent interfaces with the MCAT and the storage resources to execute a particular request. It is possible to create a federation of SRB servers by interconnecting the masters. In a federation, a server acts as a client to another server. A client request is handed over to the appropriate server depending on the location determined by the MCAT service.

SRB implements transparency for data access and transfer by managing data as collections which own and manage all of the information required for describing the data independent of the underlying storage system. The collection takes care of updating and managing consistency of the data alongwith other state information such as timestamps and audit trails. Consistency is managed by providing synchronisation mechanisms that locks stale data against access and propagates updates throughout the environment until global consistency is achieved.

SRB is one of the most widely used Data Grid technologies in various application domains around the world including the UK eScience (eDiaMoND) [44], BaBar [89], BIRN [45], IVOA [92] and the California Digital Library [129].

### 4.3.5 Other Replication Strategies

Lamehamedi, et. al [69, 120] simulate replication strategies based on the replica sites being arranged in different topologies such as ring, tree or hybrid. Each site or node maintains an index of the replicas it hosts and the other locations of these replicas that it knows. Replication of a dataset is triggered when requests for it at a site exceed some threshold. The replication strategy places a replica at a site that minimises the total access costs including both read and write costs for the datasets. The write cost considers the cost of updating all the replicas after a write at one of the replicas.

Bell, et. al [121] present an file replication strategy based on an economic model that optimises the selection of sites for creating replicas. Replication is triggered by the number of requests received for a dataset. Access mediators receive these requests and start auctions to determine the cheapest replicas. A Storage Broker participates in these auctions by offering a price at which it will sell access to a replica if it is present. If the replica is not present at the local storage element, then the broker starts an auction to replicate the requested file onto its storage if it determines that having the dataset is economically feasible. Reverse Vickrey auctions are used for the bidding process.

Lee and Weissman [122] present an architecture for dynamic replication within a service Grid. The replicas are created on the basis of each site evaluating whether its performance can be improved by requesting one more replica. The most popular services are therefore, most replicated as this will entail a performance boost by lessening the load requirements on a particular replica.

Ranganathan, et. al [123] present a dynamic replication strategy that creates copies based on trade-offs between the cost and the future benefits of creating a replica. The strategy is designed for peer-peer environments where there is a high-degree of unreliability and hence, considers a minimum number of replicas that might be required given the probability of a node being up and the accuracy of information possessed by a site in a peer-peer network.

## 4.4 Resource Allocation and Scheduling

This subsection deals with study of resource allocation and scheduling strategies within Data Grids. While Grid scheduling has been a well-researched topic, this study is limited to only those strategies that explicilty deal with transfer of data during processing. Therefore, the focus here is on features such as adapting to environments with varied data sources and scheduling jobs in order to minimise the movement of data. Table 6 summarises the scheduling strategies surveyed in this section and their classification.

Table 6: Comparison between scheduling strategies.

| Work/Project | Application Model | Scope | Data Replication | Utility Function | Locality |
|---|---|---|---|---|---|
| Casanova, et al. [130] | Bag of Tasks | Individual | Coupled | Makespan | Temporal |
| GrADS [131] | Process-level | Individual | Decoupled | Makespan | Spatial |
| Ranganathan & Foster [80] | Independent Tasks | Individual | Decoupled | Makespan | Spatial |
| Kim & Weissman [132] | Independent Tasks | Individual | Decoupled | Makespan | Spatial |
| Takefusa, et. al [133] | Process-level | Individual | Coupled | Makespan | Temporal |
| Pegasus [134] | Workflows | Individual | Decoupled | Makespan | Temporal |
| Thain, et. al [135] | Independent Tasks | Community | Coupled | Makespan | Both |
| Chameleon [136] | Independent Tasks | Individual | Decoupled | Makespan | Spatial |
| SPHINX [137][138] | Workflows | Community | Decoupled | QoS | Spatial |
| Gridbus Broker [139] & Workflow [140] | Bag of Tasks & Workflows | Individual | Decoupled | QoS | Spatial |

Casanova et. al [130] discuss heuristics for scheduling independent tasks sharing common files, on a Grid composed of interconnected clusters. The source of the files is considered to be the client host, i.e., the machine which submits the jobs to the Grid. The heuristics minimize the *makespan* or the Minimum Completion Time (MCT) of the task. Three of these heuristics: *Min-Min*, *Max-Min* and *Sufferage* were introduced by Maheswaran, et. al in [141] and are extended in this work to consider input and output data transfer times. The fourth heuristic, *XSufferage* is an extended version of Sufferage that takes into account file locality before scheduling jobs by considering MCT on the cluster level. Within XSufferage, a job is scheduled to a cluster if the file required for the job has been previously transferred to any node within the cluster. The algorithm therefore exploits the temporal locality of file requests. If a file is not present, then it is transferred over to that node which means that data transfer is coupled to job scheduling.

Scheduling within the Grid Application Development Software (GrADS) project [131] is carried out in three phases: before the execution, there is an initial matching of application requirements to available resources called *launch-time scheduling*; then, the initial schedule is modified during the execution to take into account dynamic changes in the system availability and this is called *rescheduling*; finally, the co-ordination of all schedules is done through *meta-scheduling*. The initial schedule is based on the application performance model which is matched to the resources. Contracts [142] are formed to ensure guaranteed execution performance. The rescheduling system monitors if there are any contract violations in which case it takes corrective action by migrating processes to better nodes. The metascheduler checks if the scheduling of a new application will cause any contract violations for those already scheduled. GrADS uses ClassAds [118] mechanism to specify application requirements.

The main aim of scheduling in GrADS is to reduce the execution time of the processes. The mapping and search procedure presented in [143] forms Candidate Machine Groups(CMG) consisting of available resources which are then pruned to yield one suitable group per job. The mapper then maps the application data to physical location for this group. Therefore, spatial locality is primarily exploited. The scheduler is tightly integrated into the application and works at the process level. However, the algorithms are themselves independent of the application. Recent work however has suggested extending the GrADS scheduling concept to workflow applications [144]. However, the treatment of data still remains the same.

Ranganathan and Foster [80] propose a decoupled scheduling architecture for data intensive applications which consists of 3 components: the External Scheduler (ES) that decides to which node the jobs must be submitted, the Local Scheduler (LS) on each node that decides the priority of the jobs arriving at that node and the Dataset Scheduler (DS) that tracks the popularity of the datasets and decides which datasets to replicate or delete. Through simulation, they evaluate combinations of 4 job scheduling algorithms for the ES and 3 replication algorithms for the DS. The ES schedules jobs to either a random site or the least loaded site or the site where the data is present or only the local site. The DS either does no replication or randomly replicates a file or replicates a file at the least loaded site among its neighbours. The results show that when the job is scheduled to a site where the data is available the data transfer is minimum but the response time suffers when there is no data replication. This is because a few sites which host the data are overloaded in this case and hence, making a case for dynamic replication of data.

Kim and Weissman introduce a Genetic Algorithm based scheduler for decomposable Data Grid applications in [132]. The scheduler targets an application model wherein a large dataset is split into multiple smaller datasets and these are then processed in parallel on multiple "virtual sites", where a virtual site is considered to be a collection of compute resources and data servers. The solution to the scheduling problem is represented as a chromosome whose each gene represents a task allocated to a site. Each sub-gene is associated with a value that represents the fraction of a dataset assigned to the site and the whole gene is associated with a value denoting capability of the site given the fraction of the datasets assigned, the time taken to transfer these fractions and the execution time. The chromosomes are mutated to form the next generation of chromosomes. At the end of an iteration, the chromosomes are ranked according to an objective function and the iteration stops at a predefined condition. Since the objective of the algorithm is to reduce the completion time, the iterations tend to favour those tasks in which the data is processed close to or at the point of computation thereby exploiting the spatial locality of datasets. Replication is not carried out within this algorithm and the tasks are do not have any interdependencies.

Takefusa, et. al [133] have simulated job scheduling and data replication policies for central and tier model organization of Data Grids based on the Grid Datafarm [70] architecture. The scheduling and replication policies are coupled in this evaluation to exploit locality of file access. Out of the several policies simulated, the authors establish that the combination of *OwnerComputes* strategy (job is executed on the data host that contains the data) for job scheduling alongwith background replication policies based on number of accesses(*LoadBound-Replicate*) or on the host with the maximum estimated performance (*Aggressive-Replication*) provides the minimum execution time for a job.

Workflow management in Pegasus [134] concentrates on reducing an *abstract workflow* that contains order of execution of components into a *concrete workflow* where the component is turned into an executable job and the locations of the computational resources and the data are specified. The abstract workflow goes through a process of *reduction* where the components whose outputs have already been generated and entered into a Replica Location Service are removed from the workflow and substituted with the physical location of the products. The emphasis is therefore on reuse of already produced data products. The planning process selects a source of data at random, that is, neither the temporal nor the spatial locality is exploited. The concrete workflow is then handed over to DAGMan [145] for execution.

Thain et. al [135] have described a means of creating I/O communities which are groups of CPU resources such as Condor pools clustered around a storage resource. The storage appliance satisfies the data requirements for jobs that are executed on both the processes within and outside the community. The scheduling strategy in this work allows for both the data to be staged to a community where the job is executed and for the job to migrate to a community where the data required is already staged. The decision is made by the user after comparing the overheads of either staging the application or replicating the data. The mechanism used is ClassAds alongwith indirection wherein one ClassAd refers to another ClassAd through a chain of relations specified by the user. Various different combinations of replication and scheduling were evaluated and it is found that the best performance is achieved when data local to the community is accessed by the jobs within the community. Crossing community boundaries either to move computation or data results in degradation of performance.

Chameleon [136] is a scheduler for data grid environments that takes into account the computational load of transferring the data and executables to the point of computation. They consider a data grid model consisting of sites that have computational and data resources connected by LAN and the sites are themselves interconnected by WAN. In Chameleon, therefore, a site on which the data has already been replicated is preferred over one where the data is not present to reduce the transfer time. Data replication is not coupled to scheduling in this algorithm.

SPHINX(Scheduling in Parallel for a Heterogeneous Independent NetworX) [137] is a middleware package for scheduling data-intensive applications on the Grid. Scheduling within SPHINX is based on a client-server framework in which a scheduling client within a VO submits a meta-job as a Directed Acyclic Graph (DAG) to one of the scheduling servers for the VO. The server is allocated a portion of the VO resources and in turn, it reserves some of these for the job submitted by the client and sends the client an estimate of the completion time. The server also reduces the DAG by removing tasks whose outputs are already present. If the client accepts the completion time, then the server begins execution of the reduced DAG. The scheduling strategy in SPHINX [138] considers VO policies as a four dimensional space with the resource provider, resource properties, user and time forming each of the dimensions. Each of the dimension is expressed as a hierarchy except for resource properties which is modelled as an attribute to the resource provider. Policies are therefore expressed in terms of quotas which are tuples formed by values of each dimension.The optimal resource allocation for a user request (also modelled as a tuple) is provided by a linear programming solution which minimizes the usage of the user quotas on the various resources.

Data-intensive application scheduling within the Gridbus Broker [139] is carried out on the basis of QoS factors such as deadline and budget. The execution model in this work is that of parameter sweep or Bag of Tasks, each of which depends on multiple data files each replicated on multiple data hosts. The scheduling algorithm tries to minimize the economic objective by incrementally building resource sets consisting of one compute resource for executing the job and one data site each for each file that needs to be accessed by the job. The scheduler itself performs no replication of data in this case. Scheduling of workflows is supported by the Gridbus Workflow Engine [140] which otherwise has similar properties with respect to the scheduling of data intensive applications.

# 5  Discussion and Gap Analysis

Figures 12 – 16 pictorially represent the mapping of the systems that were analysed in Section 4 to the taxonomy. Each of the boxes at the "leaves" of the taxonomy "branches" contains those systems that exhibit the property at the leaf. A box containing "(All)" implies that all the systems studied satisfy the property given by the corresponding leaf.

From the figures it can be seen that the taxonomy is shown to be complete with respect to the state-of-the-art of Data Grids as each of the systems can be fully described by the categories within this taxonomy. Some of the categories are not applicable to any of the systems that have been studied. Thus, these represent desirable properties that have not been achieved yet by the current Data Grid systems and therefore, point a way to possible future research and development in this field. In the following paragraphs, the application of the taxonomy to each of the constituent elements of the Data Grid and its implication is discussed.

Figure 12 shows the organizational taxonomy annotated with the Data Grid projects that were studied in Section 4.1. As can be seen from the figure, current scientific Data Grids mostly follow the hierarchical or the federated models of organization because the data sources are few and well-established. These data sources are generally mass storage systems from which data is transferred out as files or datasets to other repositories. From a social point of view, such Data Grids are formed by establishing collaborations between researchers from the same domain. In such cases, any new participants willing to join or contribute have to be part of the particular scientific community to be inducted into the collaboration.

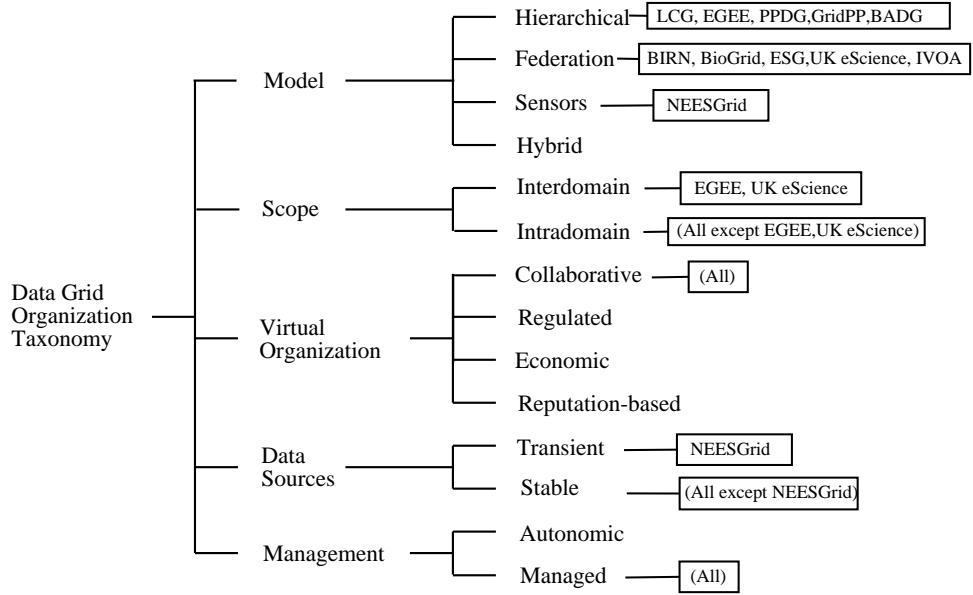

Figure 12: Mapping of Data Grid Organization Taxonomy to Data Grid Projects.

However, it is more likely that these formal structures will evolve to become more flexible in the future. This evolution will be driven by two factors. The first is the advent of sensor networks [146] as scientific and commercial tools for monitoring and measurement. While current sensor-based grids such as NEESGrid store the sensor output in a persistent storage, the demands of the future will involve more dynamic requirements such as real-time analysis of sensor data and event-triggered data capture, which in turn lead to dynamic data driven applications [147]. In such cases, Data Grids will have to deal with transient data sources. The second factor is the increasing popularity of Data Grids as a solution for large-scale computational and storage problems. This will lead to entry of commercial resource providers and therefore, will lead to market-oriented VOs wherein demand-and-supply patterns decide the price and availability of resources. This also provides incentive for content owners to offer their data for consumption outside specific domains and opens up many interesting new applications. Such VOs are likely to have a broad interdomain scope and consumers will be able to access domain-specific services by buying them off competing service providers. Also, it is more likely that these will be more autonomic as the size and scope of such VOs will make human involvement in day-to-day management difficult.

Data sources such as relational databases would become more prominent in future Data Grids. The challenge is to extend the existing Grid mechanisms such as replication, data transfer and scheduling to work with these new data sources. Work in this regard is already being done by projects such as OGSA-DAI (Data Access and Integration) [148, 149].

The mapping of various Data Grid transport mechanisms studied in Section 4.2 to the proposed taxonomy is shown in Figure 13. Data Grids require high-performance and fault-tolerant transport mechanisms. Transport protocols such as GridFTP designed for data-intensive environments help in speeding up of data transfers because of the support for parallel and striped data transfers. This also allows for maximum utilization of network links. Other features include support for end-to-end encryption of data and ability to resume data transmission from the point of disconnection.

However, data transport needs to go beyond simple movement of bits in a Data Grid. Management of the data transfers is required in order to prevent contention for network resources and to provide stronger reliability guarantees. It would also allow applications to orchestrate large data transfers so as to avoid congested or expensive links. Therefore, higher order services that build on low-level protocols are required. For example, combining GridFTP with overlay network primitives would lead to a richer set of data transfer libraries that can be used by Grid applications to manage high-speed data transfers between multiple points on the Grid. Data transfer schedulers such as Stork would then provide services to manage data transfers. Data

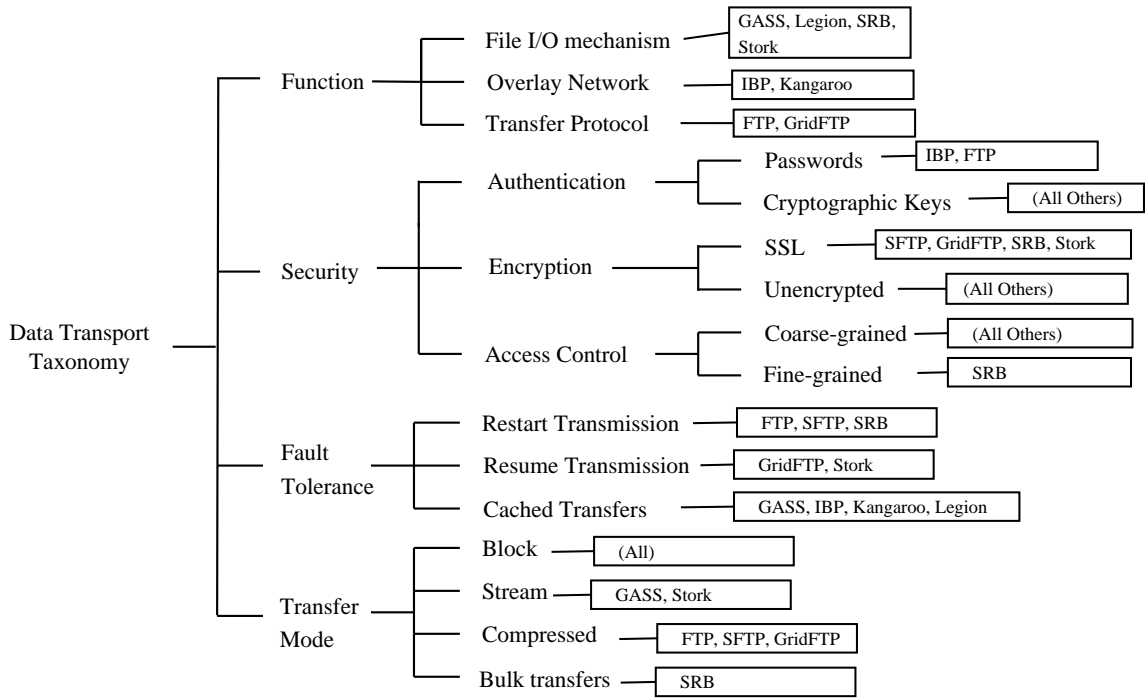

Figure 13: Mapping of Data Transport Taxonomy to Various Projects.

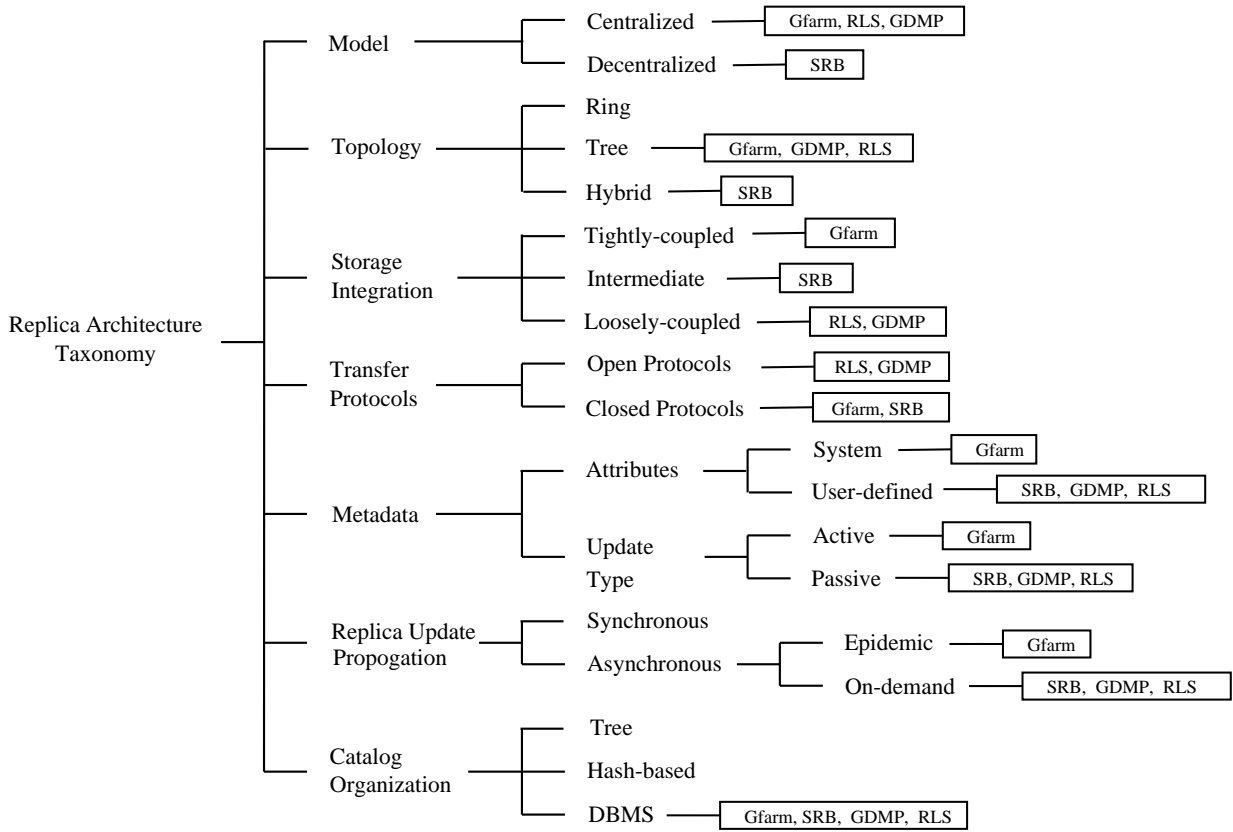

Figure 14: Mapping of Data Replication Architecture Taxonomy to Various Systems.

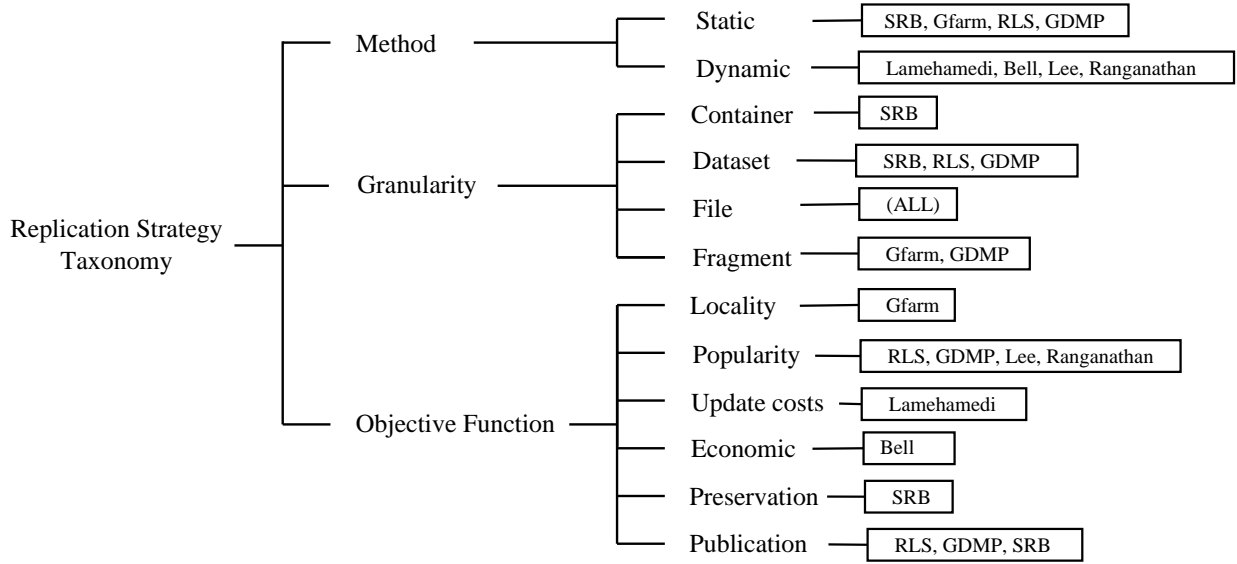

Figure 15: Mapping of Data Replication Strategy Taxonomy to Various Systems.

Grid access mechanisms also need to support fine-grained access controls such as tickets for time-limited or use-limited access and ability to lock parts of files against modifications.

Figure 14 and 15 show mapping of the data replication systems covered in Sections 4.3 to the replica architecture and strategy taxonomy. The figure shows three of the four replication mechanisms following the centralised model with a tree topology. This is no coincidence as these three were designed specifically for application to High Energy Physics domain where a hierarachical model has been widely adopted (See Section 4.1 ). Therefore, these replication mechanisms are designed to be top-down in terms of organization and data propagation. As Data Grids mature, the collaborative aspects will come to the fore leading to hybrid topologies such as trees with peer connections at the leaves.

Currently, consistency is not an issue as majority of applications deal with mostly read-only data. However, application domains such as biology and finance require stronger consistency guarantees. Additionally, consistency needs to be extended to other properties associated with data such as metadata, access controls and locks. That is change in any one of these properties of a replica must be propagated to the other replicas. Metadata is crucial for researchers in order to identify and locate the datasets or files that they require. Therefore, metadata should be bound to the data to ensure consistent updation rather than being handled separately as is the case in present replication systems. Ultimately, even procedures should be bound to the data itself so that the data becomes an active or responsive component that can describe itself and present all methods for analysis applicable to it.

From Figure 15, we gather that while dynamic replication strategies have been proposed and simulated, none have been put into action in real-world replication systems. Currently massive datasets are being replicated statically by project administrators in select locations. However, this would evolve into a more dynamic system that takes into account usage patterns of subsets of data or cost of data movement [150] before replicating. This would also be helped by application of peer-to-peer techniques such as those found in Bit-Torrent [151] in which peers simultaneously upload a file to other needy peers while downloading it from a source. The participants are encouraged to do so by making the rate of download contigent upon the rate of upload, that is, the more a client shares, the faster it receives its file. This is comparable to providing economic incentives for hosting a file.

Most of the efforts studied so far in resource allocation and scheduling in Data Grids have concentrated on reducing the makespan of the application. This can inferred from Figure 16 which illustrates mapping of scheduling efforts to the proposed taxonomy. However, focus is shifting towards providing a better Quality of Service (QoS) which is defined by the user. Within VOs, a centralised community-based scheduler would assign quotas to each of the users based on priorities and resource availability. The user scheduler should then submit jobs taking into account the assigned quotas and could negotiate with the central scheduler for quota increase or change in priorities. It could also be able to swap or reduce quotas in order to gain resource share in the future when it may be more needed. This system allows advance reservation of resources and therefore, consumers are able to plan ahead for future resource requirements. With the introduction of market-oriented VOs, there is a need for advanced application scheduling and resource allocation algorithms and policies that support optimisation of both makespan and cost of consuming grid services. New strategies are also required for efficient management of resource reservations and enhancing associated benefits. Such algorithms need to be guided by utility functions driven by profit and at the same time satisfy user-defined service quality parameters.

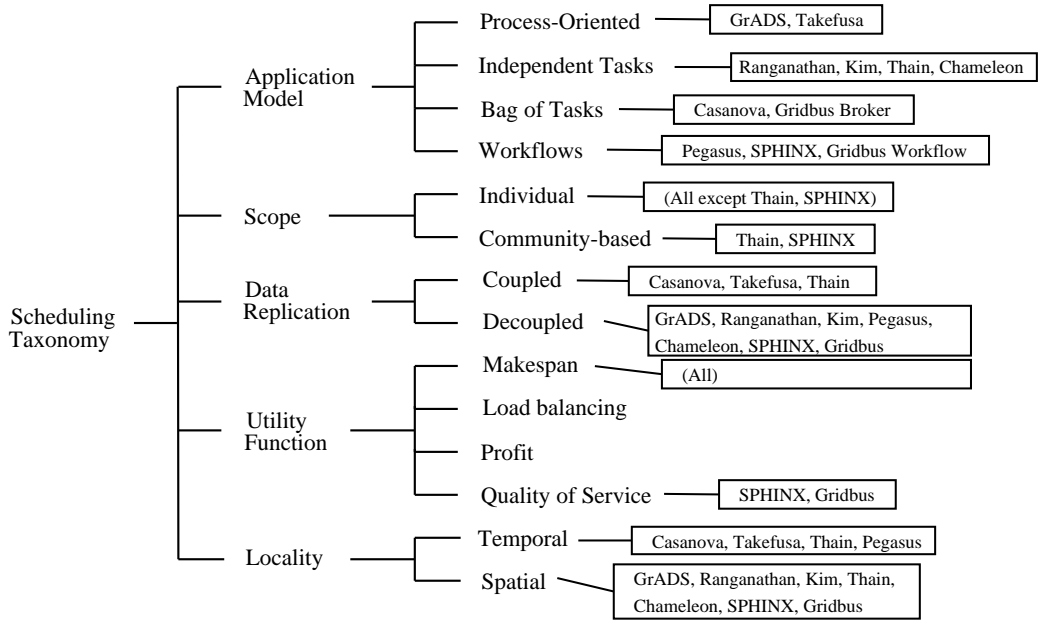

Figure 16: Mapping of Resource Allocation and Scheduling Taxonomy to Various Systems.

# 6 Summary and Conclusion

In this paper, we have studied, characterised and categorised several aspects of Data Grid systems. Data Grids have several unique features such as presence of applications with heavy computing requirements, geographically-distributed and heterogeneous resources under different administrative domains and large number of users sharing these resources and wanting to collaborate with each other. We have enumerated several characteristics where Data Grids share similarities with, and are different from, other distributed data-intensive paradigms such as content delivery networks, peer-to-peer networks and distributed databases.

Further on, we focus on the architecture of the Data Grids and the fundamental requirements of data transport mechanism, data replication systems and resource allocation and job scheduling. We develop taxonomies for each of these areas to classify the common approaches and to provide a basis for comparison of Data Grid systems and technologies. We then compared some of the representative systems in each of these areas and categorized them according to the respective taxonomies. In doing so, we have gained an insight into the architectures, strategies and practices that are currently followed within Data Grids. Also, through our characterisation, we have also been able to discover some of the shortcomings and identify gaps in the current architectures and systems. These represent some of the directions that can be followed in the future by researchers in this area. Thus, this paper lays down a comprehensive classification framework that not only serves as a tool to understanding this complex area but also presents a reference to which future efforts can be mapped.

To conclude, we have shown that Data Grids are the most appropriate platform for sharing data and col-laboratively managing and executing large-scale scientific applications that process large datasets distributed around the world. However, there more research need to be undertaken in terms of scalability, interoperability and data maintainability among others, before Data Grids can truly become the preferred platform for such applications. But, solving these problems creates the potential for Data Grids to evolve to become self-organized and self-contained and thus, creating the next generation infrastructure for enabling users to extract maximum utility out of the volumes of available information and data. A more open question is, can more sophisticated systems like relational databases and object-oriented databases be mapped to Data Grid infrastructure ?

## ACKNOWLEDGEMENTS

We would like to acknowledge the efforts of all the developers of the Grid systems surveyed in this paper. We thank our colleagues at the University of Melbourne - Krishna Nadiminti, Tianchi Ma and Sushant Goel - for their comments on this paper. We would also like to express our gratitude to Reagan Moore (San Diego Supercomputing Center) for his extensive and thought provoking comments and suggestions on various aspects of this taxonomy. We also thank Heinz Stockinger (University of Vienna), Chris Mattman (JPL, NASA) and William Allcock (Argonne National Lab) for their instructive comments on this paper. This work is partially supported through the Australian Research Council (ARC) Discovery Project grant and Storage Technology Corporation sponsorship of Grid Fellowship.

## References

[1] T. Hey and A. E. Trefethen, "The UK e-Science Core Programme and the Grid," *Journal of Future Generation Computer Systems(FGCS)*, vol. 18, no. 8, pp. 1017–1031, 2002.

[2] I. Foster and C. Kesselman, *The Grid: Blueprint for a Future Computing Infrastructure*. Morgan Kaufmann Publishers, 1999.

[3] A. Chervenak, I. Foster, C. Kesselman, C. Salisbury, and S. Tuecke, "The data grid: Towards an architecture for the distributed management and analysis of large scientific datasets," *Journal of Network and Computer Applications*, vol. 23, no. 3, pp. 187–200, 2000.

[4] P. Avery, "Data grids: a new computational infrastructure for data- intensive science," *Philosophical Transactions of the Royal Society of London Series a-Mathematical Physical and Engineering Sciences*, vol. 360, no. 1795, pp. 1191–1209, 2002.

[5] W. Hoschek, F. J. Jaen-Martinez, A. Samar, H. Stockinger, and K. Stockinger, "Data management in an international data grid project," in *Proceedings of the First IEEE/ACM International Workshop on Grid Computing(GRID '00)*. Bangalore, India: Springer-Verlag, Berlin, December 2000.

[6] K. Krauter, R. Buyya, and M. Maheswaran, "A taxonomy and survey of grid resource management systems for distributed computing," *International Journal of Software: Practice and Experience (SPE)*, vol. 32, no. 2, pp. 135–164, 2002.

[7] J. Bunn and H. Newman, *Grid Computing: Making the Global Infrastructure a Reality*. Wiley Press, UK, 2003, ch. Data Intensive Grids for High Energy Physics.

[8] X. Qin and H. Jiang, "Data Grids: Supporting Data-Intensive Applications in Wide Area Networks," University of Nebraska, Lincoln, Tech. Rep. TR-03-05-01, May 2003.

[9] R. Moore and A. Merzky, "Persistent archive basic components," Persistent Archive Research Group, Global Grid Forum, Tech. Rep., July 2002.

[10] A. Finkelstein, C. Gryce, and J. Lewis-Bowen, "Relating requirements and architectures: A study of data-grids," *Journal of Grid Computing*, vol. 2, no. 3, pp. 207–222, 2004.

[11] C. A. Mattmann, N. Medvidovic, P. Ramirez, and V. Jakobac, "Unlocking the Grid," in *Proceedings of the 8th ACM SIGSOFT Symposium on Component-based Software Engineering (CBSE8)*. St. Louis, USA: ACM Press, May 2005.

[12] R. Moore, A. Rajasekar, and M. Wan, "Data Grids, Digital Libraries and Persistent Archives: An Integrated Approach to Publishing, Sharing and Archiving Datas," *Proceedings of the IEEE (Special Issue on Grid Computing)*, vol. 93, no. 3, 2005.

[13] P. Lebrun, "The Large Hadron Collider, A Megascience Project," in *38th INFN Eloisatron Project Workshop on Superconducting Materials for High Energy Colliders*, Erice, Italy, October 1999.

[14] K. Holtman *et al.*, "Cms requirements for the grid," in *Proceedings of 2001 Conference on Computing in High Energy Physics(CHEP 2001)*. Beijing, China: Science Press, September 2001.

[15] I. Foster, C. Kesselman, and S. Tuecke, "The anatomy of the grid: Enabling scalable virtual organizations," *International Journal of High Performance Computing Applications*, vol. 15, no. 3, pp. 200–222, 2001.

[16] M. Baker, R. Buyya, and D. Laforenza, "Grids and Grid Technologies for Wide-Area Distributed Computing," *International Journal of Software: Practice and Experience (SPE)*, vol. 32, no. 15, pp. 1437–1466, December 2002, wiley Press, USA.

[17] B. D. Davison, "A web caching primer," *IEEE Internet Computing*, vol. 5, no. 4, pp. 38–45, 2001.

[18] J. Dilley, B. Maggs, J. Parikh, H. Prokop, R. Sitaraman, and B. Weihl, "Globally distributed content delivery," *IEEE Internet Computing*, vol. 6, no. 5, pp. 50– 58, 2002.

[19] B. Krishnamurthy, C. Wills, and Y. Zhang, "On the use and performance of content distribution networks," in *Proceedings of the 1st ACM SIGCOMM Workshop on Internet Measurement(IMW '01)*. San Francisco, California, USA: ACM Press, November 2001, pp. 169–182.

[20] "Akamai Inc." accessed Mar 2005. [Online]. Available: http://www.akamai.com

[21] "Speedera Inc." accessed Mar 2005. [Online]. Available: http://www.speedera.com

[22] "IntelliDNS Inc." accessed Mar 2005. [Online]. Available: http://www.intellidns.com

[23] A. Oram, *Peer-to-Peer: Harnessing the Power of Disruptive Technologies*, A. Oram, Ed. O'Reilly & Associates, Inc., 2001.

[24] D. Choon-Hoong, S. Nutanong, and R. Buyya, *Peer-to-Peer Computing: Evolution of a Disruptive Technology*. Idea Group Publishers, Hershey, PA, USA, 2005, ch. Peer-to-Peer Networks for Content Sharing, pp. 28–65.

[25] D. P. Anderson, J. Cobb, E. Korpela, M. Lebofsky, and D. Werthimer, "Seti@home: an experiment in public-resource computing," *Communications of the ACM*, vol. 45, no. 11, pp. 56–61, 2002.

[26] R. Buyya and S. Vazhkudai, "Compute Power Market: Towards a Market-Oriented Grid," in *Proceedings of the 1st International Symposium on Cluster Computing and the Grid(CCGRID '01)*. IEEE Computer Society, 2001, p. 574.

[27] "Napster Inc." [Online]. Available: http://www.napster.com

[28] "Gnutella File-Sharing Network." [Online]. Available: http://www.gnutella.com

[29] "Kazaa Inc." http://www.kazaa.com.

[30] "Jabber Protocols," accessed Mar 2005. [Online]. Available: http://www.jabber.org/protocol/

[31] D. S. Milojicic, V. Kalogeraki, R. Lukose, K. Nagaraja, J. Pruyne, B. Richard, S. Rollins, and Z. Xu, "Peer-to-peer computing," HP Labs, Palo Alto, CA, USA, Tech. Rep. HPL-2002-57, 2002.

[32] S. Ceri and G. Pelagatti, *Distributed databases : principles and systems.* McGraw-Hill, New York, 1984.

[33] M. T. Ozsu and P. Valduriez, *Principles of distributed database systems*, 2nd ed. Upper Saddle River, NJ, USA: Prentice-Hall, Inc., 1999.

[34] A. P. Sheth and J. A. Larson, "Federated database systems for managing distributed, heterogeneous, and autonomous databases," *ACM Computing Surveys*, vol. 22, no. 3, pp. 183–236, 1990.

[35] J. Gray and A. Reuter, *Transaction processing : concepts and techniques.* San Mateo, Calif: Morgan Kaufmann Publishers, 1993.

[36] I. Clarke, O. Sandberg, B. Wiley, and T. W. Hong, "Freenet: a distributed anonymous information storage and retrieval system," in *International workshop on Designing privacy enhancing technologies.* Springer-Verlag New York, Inc., 2001, pp. 46–66.

[37] I. Stoica, R. Morris, D. Liben-Nowell, D. R. Karger, M. F. Kaashoek, F. Dabek, and H. Balakrishnan, "Chord: a scalable peer-to-peer lookup protocol for internet applications," *IEEE/ACM Transactions on Networking*, vol. 11, no. 1, pp. 17–32, 2003.

[38] S. Ratnasamy, P. Francis, M. Handley, R. Karp, and S. Schenker, "A scalable content-addressable network," in *Proceedings of the 2001 conference on Applications, technologies, architectures, and protocols for computer communications(SIGCOMM '01).* ACM Press, 2001, pp. 161–172.

[39] A. I. T. Rowstron and P. Druschel, "Pastry: Scalable, decentralized object location, and routing for large-scale peer-to-peer systems," in *Proceedings of the IFIP/ACM International Conference on Distributed Systems Platforms(Middleware 2001).* Heidelberg, Germany: Springer-Verlag, 2001, pp. 329–350.

[40] B. Y. Zhao, J. D. Kubiatowicz, and A. D. Joseph, "Tapestry: An infrastructure for fault-tolerant wide-area location and," University of California at Berkeley, Tech. Rep. CSD-01-1141, 2001.

[41] E. Cohen and S. Shenker, "Replication strategies in unstructured peer-to-peer networks," in *Proceedings of the 2002 conference on Applications, technologies, architectures, and protocols for computer communications (SIGCOMM '02).* Pittsburgh, Pennsylvania, USA: ACM Press, 2002.

[42] M. Karlsson and M. Mahalingam, "Do we need replica placement algorithms in content delivery networks?" in *Proceedings of the 2002 Web Content Caching and Distribution Conference (WCW '02).* Boulder, Colorado: http://www.iwcw.org/, August 2002.

[43] B. Ciciani, D. M. Dias, and P. S. Yu, "Analysis of replication in distributed database systems," *IEEE Transactions on Knowledge and Data Engineering*, vol. 2, no. 2, pp. 247–261, 1990.

[44] "eDiaMoND Grid Computing Project," accessed Mar 2005. [Online]. Available: http://www.ediamond.ox.ac.uk/

[45] "Biomedical Informatics Research Network (BIRN)," accessed Mar 2005. [Online]. Available: http://www.nbirn.net

[46] D. Dullmann, W. Hoschek, J. Jaen-Martinez, B. Segal, A. Samar, H. Stockinger, and K. Stockinger, "Models for Replica Synchronisation and Consistency in a Data Grid," in *Proceedings of the 10th IEEE International Symposium on High Performance Distributed Computing (HPDC-10').* San Francisco, CA: IEEE Computer Society, 2001.

[47] J. Kubiatowicz, D. Bindel, Y. Chen, S. Czerwinski, P. Eaton, D. Geels, R. Gummadi, S. Rhea, H. Weatherspoon, C. Wells, and B. Zhao, "Oceanstore: an architecture for global-scale persistent storage," in *Proceedings of the ninth international conference on Architectural support for programming languages and operating systems(ASPLOS-IX).* Cambridge, Massachusetts, United States: ACM Press, 2000, pp. 190–201.

[48] J. Gray, P. Helland, P. O'Neil, and D. Shasha, "The dangers of replication and a solution," in *Proceedings of the 1996 ACM SIGMOD international conference on Management of data (SIGMOD '96)*. Montreal, Quebec, Canada: ACM Press, 1996, pp. 173–182.

[49] "Database Access and Integration Services Working Group, Global Grid Forum," accessed Mar 2005. [Online]. Available: http://www.cs.man.ac.uk/grid-db/

[50] "Transaction Management Research Group (GGF)." [Online]. Available: http://www.data-grid.org/tm-rg-charter.html

[51] R. Alonso and D. Barbara, "Negotiating data access in federated database systems," in *Proceedings of the Fifth International Conference on Data Engineering*. IEEE Computer Society, 1989, pp. 56–65.

[52] S. Saroiu, K. P. Gummadi, R. J. Dunn, S. D. Gribble, and H. M. Levy, "An analysis of internet content delivery systems," *SIGOPS Operating Systems Review*, vol. 36, no. Special Issue: Network behaviour, pp. 315–327, 2002.

[53] D. Kossmann, "The state of the art in distributed query processing," *ACM Computing Surveys*, vol. 32, no. 4, pp. 422–469, 2000.

[54] I. Foster and A. Iamnitchi, "On death, taxes, and the convergence of peer-to-peer and grid computing," in *Second International Workshop on Peer-to-Peer Systems(IPTPS)*, ser. Lecture Notes in Computer Science, vol. 2735. Berkeley, CA, USA: Springer-Verlag, Berlin, February 2003, pp. 118 – 128.

[55] J. Ledlie, J. Shneidman, M. Seltzer, and J. Huth, "Scooped, again," in *Second International Workshop on Peer-to-Peer Systems (IPTPS 2003)*, ser. Lecture Notes in Computer Science, vol. 2735. Berkeley, CA, USA,: Springer-Verlag, February 2003.

[56] G. Fox and S. Pallickara, "The narada event brokering system: Overview and extensions," in *Proceedings of the International Conference on Parallel and Distributed Processing Techniques and Applications(PDPTA '02)*. CSREA Press, 2002, pp. 353–359.

[57] M. Aderholz *et al.*, "Monarc project phase2 report," CERN, Tech. Rep., March 2000. [Online]. Available: http://monarc.web.cern.ch/MONARC/docs/phase2report/Phase2Report.pdf

[58] A. Rajasekar, M. Wan, R. Moore, and W. Schroeder, "Data Grid Federation," in *Proceedings of the 11th International Conference on Parallel and Distributed Processing Techniques and Applications (PDPTA 2004)*. Las Vegas, USA: CSREA Press, June 2004.

[59] R. Moore, A. Jagatheesan, A. Rajasekar, M. Wan, and W. Schroeder, "Data grid management systems," in *Proceedings of the 12th NASA Goddard, 21st IEEE Conference on Mass Storage Systems and Technologies*. College Park, Maryland, USA: IEEE Computer Press, April 2004.

[60] L. Pearlman, C. Kesselman, S. Gullapalli, B. Spencer Jr., J. Futrelle, R. Kathleen, I. Foster, P. Hubbard, and C. Severance, "Distributed hybrid earthquake engineering experiments: Experiences with a ground-shaking grid application," in *Proceedings of the 13th IEEE Symposium on High Performance Distributed Computing (HPDC-13)*. Honolulu, Hawaii, USA: IEEE Computer Society, Los Alamitos, CA, USA, June 2004.

[61] M. Parashar and S. Hariri, "Autonomic grid computing," in *Proceedings of the 2004 International Conference on Autonomic Computing (ICAC '04)*. New York, USA: IEEE Computer Society, May 2004, tutorial.

[62] O. Ardaiz, P. Artigas, T. Eymann, F. Freitag, L. Navarro, and R. Reinicke, "Self-organizing resource allocation for autonomic networks," in *Proceedings of the 1st International Workshop on Autonomic Computing Systems*. Prague, Czech Republic: IEEE Computer Society, September 2003.

[63] J. Postel and J. K. Reynolds, "RFC 959: File transfer protocol," Oct 1985, sTANDARD. [Online]. Available: ftp://ftp.internic.net/rfc/rfc959.txt

[64] W. Allcock, "Gridftp protocol specification," March 2003, (Global Grid Forum Recommendation GFD.20). [Online]. Available: http://www.globus.org/research/papers/GFD-R.0201.pdf

[65] D. Andersen, H. Balakrishnan, F. Kaashoek, and R. Morris, "Resilient overlay networks," in *Proceedings of the eighteenth ACM symposium on Operating systems principles(SOSP '01)*. Banff, Alberta, Canada: ACM Press, 2001, pp. 131–145.

[66] B. C. Neuman and T. Ts'o, "Kerberos: An authentication service for computer networks," *IEEE Communications*, vol. 32, no. 9, pp. 33–38, September 1994.

[67] I. Foster, C. Kesselman, G. Tsudik, and S. Tuecke, "A security architecture for computational grids," in *Proc. 5th ACM Conference on Computer and Communications Security Conference*. San Francisco, CA, USA.: ACM Press, NY, USA, November 1998.

[68] D. Wagner and B. Schneier, "Analysis of the SSL 3.0 Protocol," in *Proceedings of the Second USENIX Workshop on Electronic Commerce*. USENIX Press, November 1996.

[69] H. Lamehamedi, B. Szymanski, Z. Shentu, and E. Deelman, "Data replication strategies in grid environments," in *Proceedings of the Fifth International Conference on Algorithms and Architectures for Parallel Processing (ICA3PP'02)*, 2002.

[70] O. Tatebe, Y. Morita, S. Matsuoka, N. Soda, and S. Sekiguchi, "Grid Datafarm Architecture for Petascale Data Intensive Computing," in *2nd IEEE/ACM International Symposium on Cluster Computing and the Grid (CCGrid 2002)*. Berlin, Germany: IEEE Computer Society, May 2002.

[71] C. Baru, R. Moore, A. Rajasekar, and M. Wan, "The SDSC Storage Resource Broker," in *Procs. of CASCON'98*, Toronto, Canada, Nov 1998.

[72] A. Chervenak, E. Deelman, I. Foster, L. Guy, W. Hoschek, A. Iamnitchi, C. Kesselman, P. Kunst, M. Ripeanu, B. Schwartzkopf, H. Stockinger, K. Stockinger, and B. Tierney, "Giggle: A framework for constructing scalable replica location services," in *Proc. Supercomputing 2002 (SC 2002)*, Baltimore,USA, November 2002.

[73] A. Samar and H. Stockinger, "Grid Data Management Pilot (GDMP): A Tool for Wide Area Replication," in *Proceedings of the IASTED International Conference on Applied Informatics (AI2001)*, Innsbruck, Austria, February 2001.

[74] J. Holliday, D. Agrawal, and A. E. Abbadi, "Database replication using epidemic update," University of California at Santa Barbara, Tech. Rep. TRCS00-01, January 2000.

[75] H. Stockinger, A. Samar, B. Allcock, I. Foster, K. Holtman, and B. Tierney, "File and object replication in data grids," in *Proceedings of the 10th IEEE Symposium on High Performance and Distributed Computing (HPDC-10)*. San Francisco, USA: IEEE Computer Society Press, August 2001.

[76] B. Allcock, J. Bester, J. Bresnahan, A. Chervenak, I. Foster, C. Kesselman, S. Meder, V. Nefedova, D. Quesnel, and S. Tuecke, "Secure, efficient data transport and replica management for high-performance data-intensive computing," in *Proceedings of IEEE Mass Storage Conference*, San Diego, USA, April 2001.

[77] D. Abramson, J. Giddy, and L. Kotler, "High Performance Parametric Modeling with Nimrod/G: Killer Application for the Global Grid?" in *IPDPS'2000*, Cancun, Mexico, 2000.

[78] C. Dumitrescu and I. Foster, "Usage policy-based cpu sharing in virtual organizations," in *Proceedings of the Fifth IEEE/ACM International Workshop on Grid Computing (GRID'04)*. Pittsburgh, PA, USA: IEEE Computer Society, November 2004.

[79] G. Wasson and M. Humphrey, "Policy and enforcement in virtual organizations," in *GRID '03: Proceedings of the Fourth International Workshop on Grid Computing*. Phoenix, Arizona: IEEE Computer Society, November 2003.

[80] K. Ranganathan and I. Foster, "Decoupling Computation and Data Scheduling in Distributed Data-Intensive Applications," in *Proceedings of the 11th IEEE Symposium on High Performance Distributed Computing (HPDC)*. Edinburgh, Scotland: IEEE Computer Society, July 2002.

[81] C. D. Polychronopoulos and D. J. Kuck, "Guided self-scheduling: A practical scheduling scheme for parallel supercomputers," *IEEE Transactions on Computers*, vol. 36, no. 12, pp. 1425–1439, 1987.

[82] R. Hockauf, W. Karl, M. Leberecht, M. Oberhuber, and M. Wagner, "Exploiting spatial and temporal locality of accesses: A new hardware-based monitoring approach for dsm systems," in *Proceedings of the 4th International Euro-Par Conference on Parallel Processing(Euro-Par '98)*, ser. Lecture Notes in Computer Science, vol. 1470. Southhampton, UK: Springer-Verlag, Berlin, Germany, September 1998, pp. 206 – 215.

[83] K. S. McKinley, S. Carr, and C.-W. Tseng, "Improving data locality with loop transformations," in *ACM Trans. Program. Lang. Syst.* ACM Press, 1996, vol. 18, no. 4, pp. 424–453.

[84] A. Shatdal, C. Kant, and J. F. Naughton, "Cache conscious algorithms for relational query processing," in *Proceedings of the 20th International Conference on Very Large Data Bases(VLDB '94)*. Santiago, Chile: Morgan Kaufmann Publishers Inc., September 1994, pp. 510–521.

[85] M. Stonebraker, R. Devine, M. Kornacker, W. Litwin, A. Pfeffer, A. Sah, and C. Staelin, "An Economic Paradigm for Query Processing and Data Migration in Mariposa," in *Proceedings of 3rd International Conference on Parallel and Distributed Information Systems*. Austin, TX, USA: IEEE Computer Society, September 1994.

[86] "LHC Computing Grid," accessed Mar 2005. [Online]. Available: http://lcg.web.cern.ch/LCG/

[87] "Enabling Grids for E-SciencE (EGEE)," accessed Mar 2005. [Online]. Available: http://public.eu-egee.org/

[88] "Grid Physics Network(GriPhyN)," accessed Mar 2005. [Online]. Available: http://www.griphyn.org

[89] "Particle Physics Data Grid (PPDG)," accessed Jan 2005. [Online]. Available: http://www.ppdg.net/

[90] R. Gardner *et al.*, "The Grid2003 Production Grid: Principles and Practice," in *Proceedings of the 13th Symposium on High Performance Distributed Computing (HPDC 13)*. Honolulu, USA: IEEE CS Press, June 2004.

[91] "BioGrid, Japan," biogrid-jp. [Online]. Available: http://www.biogrid.jp/

[92] "International Virtual Observatory Alliance," accessed Mar 2005. [Online]. Available: http://www.ivoa.net/

[93] "Australian Virtual Observatory," accessed Mar 2005. [Online]. Available: http://www.aus-vo.org/

[94] B. Allcock, I. Foster, V. Nefedova, A. Chervenak, E. Deelman, C. Kesselman, J. Lee, A. Sim, A. Shoshani, B. Drach, and D. Williams, "High-performance remote access to climate simulation data: a challenge problem for data grid technologies," in *Proceedings of the 2001 ACM/IEEE conference on Supercomputing (SC '01)*. Denver, CO, USA: ACM Press, November 2001.

[95] "GridPP: UK Computing for Particle Physics," accessed Mar 2005. [Online]. Available: http://www.gridpp.ac.uk

[96] "The Belle Analysis Data Grid (BADG) Project," accessed Jan 2005. [Online]. Available: http://epp.ph.unimelb.edu.au/epp/grid/badg/about.php3

[97] "Laser Interferometer Gravitational Wave Observatory," accessed Mar 2005. [Online]. Available: http://www.ligo.caltech.edu/

[98] "Sloan Digital Sky Survey," accessed Mar 2005. [Online]. Available: http://www.sdss.org/

[99] J. Bester, I. Foster, C. Kesselman, J. Tedesco, and S. Tuecke, "GASS: A Data Movement and Access Service for Wide Area Computing Systems," in *Proceedings of the Sixth Workshop on I/O in Parallel and Distributed Systems*. Atlanta, USA: ACM Press, May 1999.

[100] J. Plank, M. Beck, W. R. Elwasif, T. Moore, M. Swany, and R. Wolski, "The Internet Backplane Protocol: Storage in the Network," in *NetStore99: The Network Storage Symposium*, Seattle, WA, USA, Oct 1999.

[101] A. Bassi, M. Beck, G. Fagg, T. Moore, J. Plank, M. Swany, and R. Wolski, "The Internet Backplane Protocol: A Study in Resource Sharing," in *2nd IEEE/ACM International Symposium on Cluster Computing and the Grid (CCGRID 2002)*, Berlin, Germany, May 2002.

[102] J. Galbraith, O. Saarenmaa, T. Ylonen, and S. Lehtinen, "SSH File Transfer Protocol (SFTP)," March 2005, internet Draft. Valid upto September 2005. [Online]. Available: http://www.ietf.org/internet-drafts/draft-ietf-secsh-filexfer-07.txt

[103] B. Allcock, J. Bester, J. Bresnahan, A. L. Chervenak, I. Foster, C. Kesselman, S. Meder, V. Nefedova, D. Quesnel, and S. Tuecke, "Data management and transfer in high-performance computational grid environments," *Parallel Computing*, vol. 28, no. 5, pp. 749–771, 2002.

[104] D. Thain, J. Basney, S.-C. Son, and M. Livny, "The Kangaroo Approach to Data Movement on the Grid," in *Proc. of the Tenth IEEE Symposium on High Performance Distributed Computing (HPDC10)*. San Francisco, California: IEEE CS Press, 2001.

[105] S. Chapin, J. Karpovich, and A. Grimshaw, "The Legion resource management system," in *Proceedings of the 5th Workshop on Job Scheduling Strategies for Parallel Processing*. IEEE Computer Society, April 1999.

[106] T. Kosar and M. Livny, "Stork: Making data placement a first class citizen in the grid," in *Proceedings of the 24th International Conference on Distributed Computing Systems (ICDCS'04)*. Tokyo, Japan: IEEE Computer Society, March 2004.

[107] K. Czajkowski, I. T. Foster, N. T. Karonis, C. Kesselman, S. Martin, W. Smith, and S. Tuecke, "A Resource Management Architecture for Metacomputing Systems," in *Proceedings of the Workshop on Job Scheduling Strategies for Parallel Processing(IPPS/SPDP '98)*. Orlando, Florida, USA: Springer-Verlag, London, UK, March 1998.

[108] J. S. Plank, T. Moore, and M. Beck, "Scalable Sharing of Wide Area Storage Resource," University of Tennessee Department of Computer Science, Tech. Rep. CS-02-475, January 2002. [Online]. Available: http://loci.cs.utk.edu/modules.php?name=Publications"&lid=109

[109] M. Horowitz and S. Lunt, "RFC 2228: FTP security extensions," oct 1997, proposed Standard. [Online]. Available: ftp://ftp.internic.net/rfc/rfc2228.txt

[110] P. Hethmon and R. Elz, "RFC 2389: Feature negotiation mechanism for the File Transfer Protocol," August 1998, proposed Standard. [Online]. Available: ftp://ftp.internic.net/rfc/rfc2389.txt

[111] I. Foster and C. Kesselman, "The Globus Project: A Status Report," in *IPPS/SPDP'98 Heterogeneous Computing Workshop*, 1998, pp. 4–18.

[112] G. von Laszewski, B. Alunkal, J. Gawor, R. Madduri, P. Plaszczak, and X.-H. Sun, "A File Transfer Component for Grids," in *Proceedings of the 2003 International Conference on Parallel and Distributed Processing Techniques and Applications*, Las Vegas, USA, June 2003.

[113] G. von Laszewski, I. Foster, J. Gawor, and P. Lane, "A Java commodity grid kit," *Concurrency and Computation-Practice and Experience*, vol. 13, no. 8-9, pp. 645–662, 2001.

[114] "NCSA GridFTP Client," accessed Mar 2005. [Online]. Available: http://dims.ncsa.uiuc.edu/set/uberftp/

[115] M. Ellert, A. Konstantinov, B. Konya, O. Smirnova, and A. Waananen, "Performance evaluation of gridftp within the nordugrid project," NorduGrid Project, Tech. Rep. cs.DC/0205023, Jan 2002. [Online]. Available: http://arxiv.org/abs/cs.DC/0205023

[116] B. S. White, A. S. Grimshaw, and A. Nguyen-Tuong, "Grid-Based File Access: The Legion I/O Model," in *Proceedings of the Ninth IEEE International Symposium on High Performance Distributed Computing (HPDC'00)*. Pittsburgh, USA: IEEE Computer Society, august 2000.

[117] A. Ferrari, F. Knabe, M. Humphrey, S. J. Chapin, and A. S. Grimshaw, "A flexible security system for metacomputing environments," in *HPCN Europe '99: Proceedings of the 7th International Conference on High-Performance Computing and Networking*. Springer-Verlag, 1999, pp. 370–380.

[118] R. Raman, M. Livny, and M. Solomon, "Matchmaking: An extensible framework for distributed resource management," *Cluster Computing*, vol. 2, no. 2, pp. 129–138, 1999.

[119] T. Kosar, G. Kola, and M. Livny, "Data pipelines: enabling large scale multi-protocol data transfers," in *Proceedings of the 2nd workshop on Middleware for grid computing(MGC '04)*. Toronto, Ontario, Canada: ACM Press, 2004.

[120] H. Lamehamedi, Z. Shentu, B. Szymanski, and E. Deelman, "Simulation of Dynamic Data Replication Strategies in Data Grids," in *Proceedings of the 17th International Symposium on Parallel and Distributed Processing(IPDPS '03)*. Nice, France: IEEE Computer Society, 2003.

[121] W. H. Bell, D. G. Cameron, R. Carvajal-Schiaffino, A. P. Millar, K. Stockinger, and F. Zini, " Evaluation of an Economy-Based File Replication Strategy for a Data Grid," in *Proceedings of the 3rd IEEE/ACM International Symposium on Cluster Computing and the Grid, 2003 (CCGrid 2003)*, Tokyo, Japan, May 2003.

[122] B.-D. Lee and J. B. Weissman, "Dynamic Replica Management in the Service Grid," in *Proceedings of the 10th IEEE International Symposium on High Performance Distributed Computing (HPDC-10')*. San Francisco, CA: IEEE Computer Society, 2001.

[123] K. Ranganathan, A. Iamnitchi, and I. Foster, "Improving data availability through dynamic model-driven replication in large peer-to-peer communities," in *2nd IEEE/ACM International Symposium on Cluster Computing and the Grid (CCGRID'02)*, Berlin, Germany, May 2002.

[124] N. Yamamoto, O. Tatebe, and S. Sekiguchi, "Parallel and Distributed Astronomical Data Analysis on Grid Datafarm," in *Proceedings of 5th IEEE/ACM International Workshop on Grid Computing (Grid 2004)*, Pittsburgh, USA, November 2004.

[125] O. Tatebe, H. Ogawa, Y. Kodama, T. Kudoh, S. Sekiguchi, S. Matsuoka, K. Aida, T. Boku, M. Sato, Y. Morita, Y. Kitatsuji, J. Williams, and J. Hicks, "The Second Trans-Pacific Grid Datafarm Testbed and Experiments for SC2003," in *Proceedings of 2004 International Symposium on Applications and the Internet - Workshops (SAINT 2004 Workshops)*, Tokyo, Japan, January 2004.

[126] O. Tatebe, N. Soda, Y. Morita, S. Matsuoka, and S. Sekiguchi, "Gfarm v2: A Grid file system that supports high-performance distributed and parallel data computing," in *2004 Computing in High Energy and Nuclear Physics (CHEP04) Conference*, Interlaken, Switzerland, September 2004.

[127] K. Czajkowski, C. Kesselman, S. Fitzgerald, and I. Foster, "Grid information services for distributed resource sharing," in *Proceedings of the 10th IEEE International Symposium on High Performance Distributed Computing (HPDC-10)*. San Francisco, California: IEEE Computer Society, August 2001.

[128] A. Rajasekar, M. Wan, R. Moore, G. Kremenek, and T. Guptil, "Data Grids, Collections, and Grid Bricks," in *Proceedings of the 20 th IEEE/11 th NASA Goddard Conference on Mass Storage Systems and Technologies (MSS'03)*. San Diego, CA, USA: IEEE Computer Society, April 2003.

[129] A. Rajasekar, R. Moore, B. Ludascher, and I. Zaslavsky, "The GRID Adventures: SDSC'S Storage Resource Broker and Web Services in Digital Library Applications," in *Proceedings of the Fourth All-Russian Scientific Conference (RCDL'02) Digital Libraries: Advanced Methods and Technologies, Digital Collections*, 2002.

[130] H. Casanova, A. Legrand, D. Zagorodnov, and F. Berman, "Heuristics for Scheduling Parameter Sweep Applications in Grid environments," in *9th Heterogeneous Computing Systems Workshop (HCW 2000)*. Cancun,Mexico: IEEE CS Press, 2000.

[131] H. Dail, O. Sievert, F. Berman, H. Casanova, A. YarKhan, S. Vadhiyar, J. Dongarra, C. Liu, L. Yang, D. Angulo, and I. Foster, *Grid resource management: state of the art and future trends*. Kluwer Academic Publishers, MA, USA, 2004, ch. Scheduling in the Grid application development software project, pp. 73–98.

[132] S. Kim and J. Weissman, "A GA-based Approach for Scheduling Decomposable Data Grid Applications," in *Proceedings of the 2004 International Conference on Parallel Processing (ICPP 04)*. Montreal, Canada: IEEE CS Press, August 2003.

[133] A. Takefusa, O. Tatebe, S. Matsuoka, and Y. Morita, "Performance Analysis of Scheduling and Replication Algorithms on Grid Datafarm Architecture for High-Energy Physics Applications," in *Proceedings of the 12th IEEE international Symposium on High Performance Distributed Computing(HPDC-12)*. Seattle, USA: IEEE CS Press, June 2003.

[134] E. Deelman, J. Blythe, Y. Gil, and C. Kesselman, *Grid Resource Management: State of the Art and Future Trends*. Kluwer Academic Publishers, 2003, ch. Workflow Management in GriPhyN, pp. 99–117.

[135] D. Thain, J. Bent, A. Arpaci-Dusseau, R. Arpaci-Dusseau, and M. Livny, "Gathering at the well: Creating communities for grid I/O," in *Proceedings of Supercomputing 2001*. Denver, Colorado: IEEE CS Press, November 2001.

[136] S.-M. Park and J.-H. Kim, "Chameleon: A Resource Scheduler in a Data Grid Environment," in *Proceedings of the 3rd IEEE/ACM International Symposium on Cluster Computing and the Grid, 2003 (CCGrid 2003)*. Tokyo, Japan: IEEE CS Press, May 2003.

[137] J.-U. In, A. Arbree, P. Avery, R. Cavanaugh, S. Katageri, and S. Ranka, "Sphinx: A scheduling middleware for data intensive applications on a grid," GriPhyN(Grid Physics Network), Tech. Rep. GriPhyN 2003-17, May 2003.

[138] J.-U. In, P. Avery, R. Cavanaugh, and S. Ranka, "Policy based scheduling for simple quality of service in grid computing," in *Proceedings of the 18th International Parallel and Distributed Processing Symposium 2004(IPDPS '04)*. Santa Fe, NM: IEEE Computer Society Press, April 2004.

[139] S. Venugopal and R. Buyya, "An economy-based algorithm for scheduling data-intensive applications on global grids," Grid Computing and Distributed Systems Laboratory, University of Melbourne, Australia, Tech. Rep. GRIDS-TR-2004-11, Dec 2004.

[140] J. Yu and R. Buyya, "A novel architecture for realizing grid workflow using tuple spaces," in *Proceedings of the Fifth IEEE/ACM International Workshop on Grid Computing (GRID'04)*. Pittsburgh, PA, USA: IEEE Computer Society, November 2004.

[141] M. Maheswaran, S. Ali, H. J. Siegel, D. Hensgen, and R. F. Freund, "Dynamic Mapping of a Class of Independent Tasks onto Heterogeneous Computing Systems," *Journal of Parallel and Distributed Computing(JPDC)*, vol. 59, pp. 107–131, Nov 1999.

[142] F. Vraalsen, R. Aydt, C. Mendes, and D. Reed, "Performance contracts: Predicting and monitoring grid application behavior," in *Proceedings of the 2nd International Workshop on Grid Computing(GRID 2001)*, ser. Lecture Notes in Computer Science, vol. 2242. Denver, CO: Springer-Verlag, November 2001.

[143] H. Dail, H. Casanova, and F. Berman, "A Decoupled Scheduling Approach for the GrADS Environment," in *Proceedings of the 2002 IEEE/ACM Conference on Supercomputing (SC'02)*. Baltimore, USA: IEEE CS Press, November 2002.

[144] K. Cooper, A. Dasgupata, K. Kennedy, C. Koelbel, A. Mandal, G. Marin, M. Mazina, J. Mellor-Crummey, F. Berman, H. Casanova, A. Chien, H. Dail, X. Liu, A. Olugbile, O. Sievert, H. Xia, L. Johnsson, B. Liu, M. Patel, D. Reed, W. Deng, C. Mendes, Z. Shi, A. YarKhan, and J. Dongarra, "New grid scheduling and rescheduling methods in the grads project," in *Proceedings of NSF Next Generation Software Workshop:International Parallel and Distributed Processing Symposium*. Santa Fe, USA: IEEE CS Press, April 2004.

[145] "DAGMan (Directed Acyclic Graph Manager)," accessed Mar 2005. [Online]. Available: http://www.cs.wisc.edu/condor/dagman/

[146] I. Akyildiz, W. Su, Y. Sankarasubramaniam, and E. Cayirci, "A survey on sensor networks," *IEEE Communications Magazine*, vol. 40, pp. 102–114, 2002.

[147] F. Darema, "Grid computing and beyond: the context of dynamic data driven applications systems," *Proceedings of the IEEE*, vol. 93, no. 3, pp. 692– 697, March 2005.

[148] J. Magowan, "A view on relational data on the Grid," in *Proceedings of the 17th International Symposium on Parallel and Distributed Processing(IPDPS '03)*. Nice, France: IEEE Computer Society, 2003.

[149] S. Malaika, A. Eisenberg, and J. Melton, "Standards for databases on the grid," *SIGMOD Rec.*, vol. 32, no. 3, pp. 92–100, 2003.

[150] H. Lin, "Economy-Based Data Replication Broker Policies in Data Grids," University of Melbourne, Australia, Tech. Rep., Jan 2005, bSc Honours Thesis. [Online]. Available: http://www.gridbus.org/students/HenrẏHonsThesis2004.pdf

[151] B. Cohen, "Incentives build robustness in BitTorrent," in *Proceedings of 1st Workshop on Economics of Peer-to-Peer Systems*, Berkeley, CA, June 2003.